\documentclass[smallextended,10pt,psfig,a4]{amsart}
\usepackage{geometry}
\usepackage{graphics}
\usepackage[dvips]{graphicx}

\newtheorem{theorem}{Theorem}[section]

\newtheorem{pro}[theorem]{Proposition}
\newtheorem{cor}[theorem]{Corollary}
\newtheorem{remark}[theorem]{Remark}

\theoremstyle{definition}
\newtheorem{definition}[theorem]{Definition}
\newtheorem{example}[theorem]{Example}

\theoremstyle{remark}

\numberwithin{equation}{section}

\begin{document}
\pagestyle{plain}

\author{Carlos F. Lardizabal and Rafael R. Souza}

\address{Instituto de Matem\'atica - Universidade Federal do Rio Grande do Sul - UFRGS - Av. Bento Gon\c calves 9500 - CEP 91509-900 Porto Alegre, RS, Brazil}
\email{cfelipe@mat.ufrgs.br, rafars@mat.ufrgs.br}

\def\laa{\langle}
\def\raa{\rangle}
\def\qed{\begin{flushright} $\square$ \end{flushright}}
\def\qee{\begin{flushright} $\Diamond$ \end{flushright}}
\def\ov{\overline}

\begin{abstract}
In this work we study certain aspects of Open Quantum Random Walks (OQRWs), a class of quantum channels described by S. Attal et al. \cite{attal}. As a first objective we consider processes which are nonhomogeneous in time, i.e., at each time step, a possibly distinct evolution kernel. Inspired by a spectral technique described by L. Saloff-Coste and J. Z\'u\~niga \cite{saloff}, we define a notion of ergodicity for finite nonhomogeneous quantum Markov chains and describe a criterion for ergodicity of such objects in terms of singular values. As a second objective, and based on a quantum trajectory approach, we study a notion of hitting time for OQRWs and we see that many constructions are variations of well-known classical probability results, with the density matrix degree of freedom on each site giving rise to systems which are seen to be nonclassical. In this way we are able to examine open quantum versions of the gambler's ruin, birth-and-death chain and a basic theorem on potential theory.
\end{abstract}

\date{\today}
\title{Open quantum random walks: ergodicity, hitting times, gambler's ruin and potential theory}

\maketitle

\tableofcontents

\section{Introduction}

The study of asymptotic behavior of trace-preserving completely positive maps, also known as quantum channels, is a fundamental topic in quantum information theory, see for instance \cite{burgarth,burgarth2,lardi,petulante,petulante2,novotny,novotny2}. More recently, an important class of quantum channels, namely Open Quantum Random Walks (OQRWs) has been introduced by S. Attal et al. \cite{attal} and its long term behavior studied \cite{attal2,konno,cfrr,pellegrini}. OQRWs are such that probability calculations can be expressed in terms of a trace functional (on a noncommutative domain) but their asymptotic limits are seen to present a classical character. A channel of this kind can be written as
\begin{equation}
\Phi(\rho)=\sum_{i}\Big(\sum_j B_{j}^i\rho_j B_{j}^{i*}\Big)\otimes|i\rangle\langle i|,\;\;\;\rho=\sum_i\rho_i\otimes|i\rangle\langle i|, \;\;\;\sum_i tr(\rho_i)=1,
\end{equation}
where the $B_{j}^i$ are matrices associated to a transition from site $|j\rangle$ to site $|i\rangle$, for all $i,j$ \cite{attal}. We also impose the condition $\sum_i B_{j}^{i*}B_{j}^i=I$ for all $j$. In recent works R. Carbone and Y. Pautrat \cite{carbone,carbone2} have studied irreducibility and periodicity aspects of OQRWs. As expected, the dynamical behavior of an OQRW is in general quite different from what is obtained with the usual (closed) quantum random walk \cite{portugal,salvador}, but see section 10 of \cite{attal} for a relation between both kinds.

\medskip

OQRWs are closely related to Transition Effect Matrices (TEMs), presented by S. Gudder \cite{gudder2,gudder}, and this relation (also mentioned in \cite{attal}) is a starting point for the discussion of this work. TEMs are a generalization of stochastic matrices and a typical example, acting on two sites, can be written as
\begin{equation}\label{twosites}
B(S)=\begin{bmatrix} B_{1}^1 & B_{2}^1 \\ B_{1}^2 & B_{2}^2 \end{bmatrix}\begin{bmatrix} S_1 \\ S_2 \end{bmatrix}:=\begin{bmatrix} B_{1}^1\circ S_1 + B_{2}^1\circ S_2 \\ B_{1}^2\circ S_1 + B_{2}^2\circ S_2 \end{bmatrix},\;\;\; S=\begin{bmatrix} S_1 \\ S_2\end{bmatrix},\;\;\; tr(S_1)+tr(S_2)=1,
\end{equation}
where the  positive semidefinite matrices $0\leq B_i^j\leq I$ are called effect matrices, $\sum_j B_{i}^j=I$ for all $i$, $S_i\geq 0$ and $A\circ B$ is the sequential product $A^{1/2}BA^{1/2}$.

\medskip

Instead of the conditions $0\leq B_i^j\leq I$, $\sum_j B_{i}^j=I$, one may ask what happens if we allow any matrices $B_{i}^j$ satisfying $\sum_{j} B_{i}^{j*}B_{i}^j=I$, all $i$, and replace the sequential product by the operation $(A,B)\mapsto ABA^*$. The object $B=(B_{i}^j)$ will then be called a {\bf Quantum Transition Matrix (QTM)} and we obtain a natural correspondence between such matrix and an OQRW (to be reviewed later). Then we may consider a sequence of QTMs $\mathcal{B}=\{B_1,B_2,\dots\}$, which we call a time-nonhomogeneous {\bf Quantum Markov Chain (QMC)}, and study its long term behavior. In terms of classical probability, this corresponds to the problem of studying a time-nonhomogeneous Markov chain: at each (discrete) time step, a (possibly) distinct stochastic kernel \cite{bremaud}. In our setting this means that we may consider at each time step a distinct quantum channel and we would like to describe facts about its asymptotic limit.

\medskip

Let $\mathcal{H}, \mathcal{K}$ denote Hilbert spaces, $B(\mathcal{H}\otimes\mathcal{K})$ denote the bounded operators and $D(\mathcal{H}\otimes\mathcal{K})$ the states (density matrices) on $\mathcal{H}\otimes\mathcal{K}$. By repeatedly applying an OQRW $\Phi:B(\mathcal{H}\otimes\mathcal{K})\to B(\mathcal{H}\otimes\mathcal{K})$ and a measurement of the position (projection on $\mathcal{K}$) we obtain a sequence of states on $\mathcal{H}\otimes\mathcal{K}$: if the state of the chain at time $n$ is $\rho^{(n)}=\rho\otimes|j\rangle\langle j|$ then at time $n+1$ it jumps to one of the values
\begin{equation}
\rho^{(n+1)}=\frac{B_j^i\rho B_j^{i*}}{p(j,i)}\otimes |i\rangle\langle i|,\;\;\;i\in\mathbb{Z}
\end{equation}
with probability
\begin{equation}\label{basprobfor}
p(j,i)=tr(B_j^i\rho B_j^{i*})
\end{equation}
This is a well-known transition rule seen in quantum mechanics, which depends on a density matrix. A similar description is possible if the initial state is pure and in this case the system stays valued in pure states. In probability notation, we have a homogeneous Markov chain $(\rho_n,X_n)$ with values in $D(\mathcal{H})\times \mathbb{Z}$ (continuous state, discrete time) satisfying: from any position $(\rho,j)$ one jumps to
\begin{equation}\label{qtapp}
\Bigg(\frac{B_j^i\rho B_j^{i*}}{p(j,i)},j\Bigg)
\end{equation}
with probability given by (\ref{basprobfor}). This the quantum trajectories approach of OQRWs. Also see \cite{attal,attal2,maassen}. This provides a convenient probability formalism for the problems we wish to study here. We also refer the reader to \cite{bllt2} where related calculations on quantum stochastic processes are made.

\medskip

This work has two main goals, both being applications to OQRWs:

\begin{enumerate}
\item We define a notion of ergodicity for sequences of (in general distinct) finite QTMs and prove an ergodicity criterion in terms of singular values. In this setting we have a nonhomogeneous aspect: at each time step, we apply a distinct quantum channel. It is seen that the result is somewhat expected and we will restrict ourselves to sequences of unital (unit preserving) QTMs acting on a finite number of sites. The study of ergodicity is a basic objective in the theory of Markovian dynamics and similar models, and for quantum Markov processes one already has a good understanding of ergodicity and mixing times in operator algebra contexts \cite{szehr,temme}. However, results on ergodicity of sequences of quantum channels besides the case of taking iterations of a single channel are scarce and we hope the description presented here for open quantum walks will encourage further studies. The presentation given here serves as an illustration of how a classical technique can be translated to a quantum setting.

    \smallskip

\item We use the quantum trajectories formalism to define a notion of hitting time (first visit) for OQRWs. In this setting we note that transition probabilities depend on density matrices and these are seen to vary with time. Above we noted that quantum trajectories can be studied in terms of a Markov chain with values on $D(\mathcal{H})\times \mathbb{Z}$. Another way of visualizing this process is to simply consider the state space to be $\mathbb{Z}$ but with varying probabilities (all depending on a fixed initial density). So we are interested in a kind of first visit of sites (space $\mathcal{K}$). We believe this is a point of view which is convenient for the study of certain applications.
\end{enumerate}

As a consequence of the first goal, we are able to examine the following problem. We recall from classical probability theory that a finite, irreducible, aperiodic Markov chain is such that the columns (rows) of the iterates of the associated column (row) stochastic matrix $P^r$ converge to the unique invariant distribution vector for $P$, as $r\to\infty$ \cite{norris}. In this work we prove an analogous property for QTMs satisfying certain conditions.

\medskip

In Section \ref{sec2} we establish basic terminology and notations, review OQRWs and its correspondence with QTMs. In Section \ref{sec3}, following \cite{carbone, carbone2}, we briefly discuss irreducibility and periodicity of channels, as this will serve as a review and as a motivation for our discussion on ergodicity of sequences of QTMs. Section \ref{sec5} presents a notion of ergodicity for sequences of QTMs (Definition \ref{ergodef}) and the spectral construction needed for one of our our main results. This construction is closely inspired by results presented by L. Saloff-Coste and J. Z\'u\~niga  \cite{saloff} in a classical setting, concerning time-nonhomogeneous Markov chains on a finite state space. We also relate the notion of ergodicity presented here with weak ergodicity seen in a noncommutative $L^1$-space setting \cite{mukhamedov}.

\medskip

Inspired by the transition rules obtained from quantum trajectories (depending on density matrices), we define a notion of hitting time in the setting of OQRWs (Section \ref{zxsec3}). We prove results which generalize certain Markov chains theorems and we give special attention to a minimality result, as this is also associated to questions on potential theory. Some of the proofs on this matter are closely inspired by classical descriptions, such as the ones presented by J. Norris \cite{norris} and also by P. Br\'emaud \cite{bremaud}. Some analogies are quite strong for the case of OQRWs, since these are quantum channels on the space $\mathcal{H}\otimes\mathcal{K}$, where $\mathcal{K}$ has the interpretation of being the space of sites that the walk may reach.

\medskip

Denote by $h_i^A(\rho)$ the probability that the walk will ever reach set $A$, given that the walk started at site $i$, with initial density matrix $\rho\otimes |i\rangle\langle i|$. We prove:

\medskip

{\bf Theorem \ref{minimality_th}.}
For every density $\rho\otimes |i\rangle\langle i|$, the vector of hitting probabilities $(h_i^A(\rho))_{i\in\mathbb{Z}}$ is the minimal nonnegative solution to the system
\begin{equation}\label{qver_hit1}
\left\{
\begin{array}{ll}
x_i(\rho)=1 & \textrm{ if } i\in A  \\
\sum_j tr(B_i^j\rho B_i^{j*})x_j\Big(\frac{B_i^j\rho B_i^{j*}}{tr(B_i^j\rho B_i^{j*})}\Big)=x_i(\rho)
 & \textrm{ if } i\notin A
\end{array} \right.
\end{equation}
Minimality means that if $x=(x_i)_{i\in\mathbb{Z}}$, is another solution with $x_i(\rho)\geq 0$ for all $i$ then $x_i(\rho)\geq h_i^A(\rho)$ for all $i$. Solutions are assumed to be linear functionals acting on the space of density matrices $D(\mathcal{H}\otimes\mathcal{K})$ associated to the given OQRW $\Phi$.

\medskip

The above theorem will be proved in Section \ref{zxsec4}, inspired by classical results. We call the second equation in (\ref{qver_hit1}) the {\bf open quantum version} of $Px=x$, which is the matrix equation corresponding to the classical first visit problem.  The proof goes through the fact that the stochastic matrix $P$ from the classical setting should give way to transition probabilities that depend on density matrices, and these can be seen to change with time. Nevertheless, we will see that the adaptations are based on simple additional ideas.

\medskip

We also study a particular class of channels, namely, nearest neighbor OQRWs induced by normal commuting matrices, thus making clear that we have examples in a family which is strictly larger than the set of real stochastic matrices (Section \ref{zxsec5}). For different, but related contexts where commuting contractions have been an object of study, see \cite{lmei,paulsen}. In this form we are able to study open quantum versions of the gambler's ruin and the birth-and-death chain in Section \ref{zxsec6}. As a result we will obtain outcomes which are generalizations of classical results and this is essentially due to the density matrix degree of freedom on each site of the walk. Also a simple adaptation of previous results provides a version of the average cost theorem from classical potential theory.

\medskip

{\bf Theorem \ref{teo_pot11}.}
Suppose that $\{c(i):i\in D\}$ and $\{f(i):i\in\partial D\}$ are nonnegative. Let $T$ be the hitting time of $\partial D$ and define
\begin{equation}
\phi_i(\rho):=\sum_{X_0=i,X_1,X_2,\dots\in(-a,a), X_r\in\{-a,a\}}[c(X_0)+c(X_1)+c(X_2)+\cdots+c(X_{r-1})+f(X_r)]P(X_1,\dots,X_r;\rho)
\end{equation}
Then for every density $\rho$ the following holds. a) The potential $\phi=\{\phi_i:i\in I\}$ satisfies
\begin{equation}\label{cond_itema0}
\phi_i(\rho)=c(i)+\sum_{j\in I}p_{ij}(\rho)\phi_i(\rho|X_1=j)=c(i)+\sum_{j\in I}tr(B_i^j\rho B_i^{j*})\phi_j\Big(\frac{B_i^j\rho B_i^{j*}}{tr(B_i^j\rho B_i^{j*})}\Big),
\end{equation}
and $\phi=f$ in $\partial D$. b) If $\{\psi_i:i\in I\}$ satisfies the open quantum version of $\psi\geq c+P\psi$ in $D$ and $\psi\geq f$ in $\partial D$ and $\psi_i\geq 0$ for all $i$ then $\psi_i\geq \phi_i$ for all $i$. c) If $P_i(T<\infty)=1$ for all $i$ then (\ref{cond_itema0})  has at most one bounded solution.

\medskip

We call (\ref{cond_itema0}) the {\bf open quantum version} of $\phi=c+P\phi$, which is the matrix equation corresponding to the classical potential problem. For more on potential theory with a quantum context in view, see \cite{konno2}. We conclude with a brief discussion on hitting times for nonhomogeneous quantum Markov chains and open questions.

\begin{remark} This work can be seen as the combination of two projects, one treating nonhomogeneous quantum Markov chains and another describing hitting times in an open quantum context. Besides the evident quantum motivation, these programs have in common the setting of OQRWs and certain nonhomogeneous aspects in the form of time-changing probability calculations. We hope that each discussion will enrich the other and describe a nontrivial setting where many problems can be discussed (see the open questions section at the end of the work).
\end{remark}

\medskip

{\bf Acknowledgments.} The authors are grateful to an anonymous referee for several useful suggestions that led to marked improvements of the paper. C.F.L. is partially supported by a CAPES/PROAP grant to the Graduate Program in Mathematics - PPGMat/UFRGS. R.R.S. is partially supported by FAPERGS (proc. 002063-2551/13-0). The authors would like to thank C. Liu, T. Machida, S. E. Venegas-Andraca, N. Petulante, F. Petruccione and F. A. Gr\"unbaum for stimulating discussions on this line of research.

\section{Preliminaries: Quantum Transition Matrices and Open Quantum Random Walks}\label{sec2}

In order to describe the correspondence between QTMs and OQRWs, we review basic facts about completely positive maps, more details of which can be seen for instance in \cite{alicki,benatti,nielsen,petz,watrous,wolf}. Let $\Phi: M_n(\mathbb{C})\to M_n(\mathbb{C})$ be linear. We say $\Phi$ is a {\bf positive} operator whenever $A\geq 0$ implies $\Phi(A)\geq 0$. Define for each $k\geq 1$, $\Phi_k: M_k(M_n(\mathbb{C}))\to M_k(M_n(\mathbb{C}))$,
\begin{equation}
\Phi_k(A)=[\Phi(A_{ij})],\;\;\; A\in M_k(M_n(\mathbb{C})), \; A_{ij}\in M_n(\mathbb{C})
\end{equation}
We say $\Phi$ is {\bf $k$-positive} if $\Phi_k$ is positive, and we say $\Phi$ is completely positive (CP) if $\Phi_k$ is positive for every $k=1, 2, 3,\dots$. It is well-known that CP maps can be written in the Kraus form \cite{petz}:
\begin{equation}\label{krausform}
\Phi(\rho)=\sum_i A_i\rho A_i^*
\end{equation}
We say $\Phi$ is {\bf trace-preserving} if $tr(\Phi(\rho))=tr(\rho)$ for all $\rho\in M_d(\mathbb{C})$, which is equivalent to $\sum_i A_i^*A_i=I$. We say $\Phi$ is {\bf unital} if $\Phi(I)=I$, which is equivalent to $\sum_i A_iA_i^*=I$. Trace-preserving completely positive (CPT) maps are also called {\bf quantum channels}. Also recall that if $A\in M_d(\mathbb{C})$ there is the corresponding vector representation $vec(A)$ associated to it, given by stacking together the matrix rows. For instance, if $d=2$,
\begin{equation}
A=\begin{bmatrix} a_{11} & a_{12} \\ a_{21} & a_{22}\end{bmatrix}\;\;\;\Rightarrow \;\;\; vec(A)=\begin{bmatrix} a_{11} & a_{12} & a_{21} & a_{22}\end{bmatrix}^T.
\end{equation}
The $vec$ mapping satisfies $vec(AXB^T)=(A\otimes B)vec(X)$ for any $A, B, X$ square matrices \cite{hj2} so in particular, $vec(AXA^*)=vec(AX\ov{A}^T)=(A\otimes \ov{A})vec(X)$,
from which we can obtain the {\bf matrix representation} $[\Phi]$ for the CP map (\ref{krausform}):
\begin{equation}\label{matrep}
[\Phi]=\sum_{i} A_{i}\otimes \ov{A_{i}}=\sum_{i,j,k,l}\langle E_{kl},\Phi(E_{ij})\rangle E_{ki}\otimes E_{lj}
\end{equation}
We recall the well-known fact that the matrix representation of a CPT map $\Phi: M_d(\mathbb{C})\to M_d(\mathbb{C})$ is independent of the Kraus representation considered. The proof of this result is a simple consequence of the unitary equivalence of Kraus matrices for a given quantum channel \cite{petz}.

\subsection{Correspondence between QTMs and OQRWs}\label{bigrem}

Let $\mathcal{K}$ denote a separable Hilbert space and let $\{|i\rangle\}_{i\in \mathbb{Z}}$ be an orthonormal basis for such space (in case $\mathcal{K}$ is infinite dimensional). The elements of such basis will be called sites (or vertices). Let $\mathcal{H}$ be another Hilbert space, which will describe the degrees of freedom given at each point of $\mathbb{Z}$. Then we will consider the space $\mathcal{H}\otimes\mathcal{K}$. For each pair $i,j$ we associate a bounded operator $B_{i}^j$ on $\mathcal{H}$. This operator describes the effect of passing from $|i\rangle$ to $|j\rangle$. We will assume that for each $i$, $\sum_j B_{i}^{j*}B_{i}^j=I$, where, if infinite, such series is strongly convergent. This constraint means: the sum of all the effects leaving site $i$ is $I$. We will consider density matrices on $\mathcal{H}\otimes\mathcal{K}$ with the particular form $\rho=\sum_i\rho_i\otimes |i\rangle\langle i|$, assuming that $\sum_i tr(\rho_i)=1$ (see Remark \ref{bigrem2} below).
For a given initial state of such form, the OQRW induced by the $B_{i}^j$ is, by definition,
\begin{equation}\label{oqrwdefinition}
\Phi(\rho)=\sum_i\Big(\sum_j B_{j}^i\rho_j B_{j}^{i*}\Big)\otimes |i\rangle\langle i|.
\end{equation}
We say the OQRW is {\bf finite} if $\dim(\mathcal{H})<\infty$ and if it acts on a finite number of sites, that is, $\dim(\mathcal{K})<\infty$. Now let $B=(B_{i}^j)$ denote a QTM (recall the definition following eq. (\ref{twosites}) in the Introduction). If $S=(S_i)$ is a vector state then it is clear that the calculation for $B(S)$, defined by
\begin{equation}\label{ac11}
B(S)=\begin{bmatrix} B_{1}^1 & B_{2}^1 & \cdots \\ B_{1}^2 & B_{2}^2 & \cdots \\ \vdots & \vdots & \vdots \end{bmatrix}\cdot\begin{bmatrix} S_1 \\ S_2 \\\vdots \end{bmatrix}:=\begin{bmatrix} \sum_j B_{j}^1S_jB_{j}^{1*}\\ \sum_j B_{j}^2S_jB_{j}^{2*}\\ \vdots  \end{bmatrix}
\end{equation}
can be identified with the channel iteration
\begin{equation}\label{tens_is_dirsum}
\Phi_B(S)=\sum_i\Big(\sum_j B_{j}^i S_j B_{j}^{i*}\Big)\otimes |i\rangle\langle i|
\end{equation}
Comparing (\ref{ac11}) and (\ref{tens_is_dirsum}) we realize that the projections $|i\rangle\langle i|$ are being used simply as an index in the vector state space. We conclude that a QTM induces an OQRW in a natural way and conversely.

\medskip

As a conclusion, in this work we will state results in terms of QTMs but, based by the above description, this is equivalent to making use of an associated OQRWs whenever one feels it is convenient. Given a QTM $B$, we may consider the associated OQRW $\Phi_B$ in its operator form, or still, its matrix representation $[\Phi_B]$. In other words, in order to perform calculations or study properties of a QTM $B$, it is often enough to understand the behavior of the matrix representation of the quantum channel induced by $B$.

\begin{remark}\label{bigrem2}
Even though $\Phi_B$ given by (\ref{tens_is_dirsum}) acts on a tensor product space, the resulting dynamics occurs on a direct sum space. More precisely, whatever is the initial state $\rho$ on $\mathcal{H}\otimes\mathcal{K}$, the density $\Phi_B(\rho)$ produced is of the form $\sum_i\rho_i\otimes |i\rangle\langle i|$, see \cite{attal}.
\end{remark}

\section{Reducibility and periodicity of OQRWs and QTMs}\label{sec3}

We review some facts from \cite{carbone,carbone2} and, in this section and the next, all QTMs are assumed to be finite. Following the definitions of irreducibility and periodicity, seen in classical probability and in operator algebra settings, the mentioned authors study the corresponding notions for quantum channels. In the end of this section we state the problem we wish to solve, motivated by the theory discussed here. We let $I_1(\mathcal{H})$ be the ideal of trace operators on some given Hilbert space $\mathcal{H}$.
A positive map $\Phi$ is called {\bf irreducible} if the only orthogonal projections $P$ such that
$\Phi(PI_1(\mathcal{H})P)\subset PI_1(\mathcal{H})P$ are $P=0$ and $P=Id$. When the above expression holds we say that $P$ {\bf reduces} $\Phi$. In case $\Phi$ is CP, we can characterize irreducibility in terms of its operator sum representation. We also recall the following.
\begin{definition}
Let $\Phi$ be a positive trace-preserving irreducible map and let $(P_0,\dots,P_{d-1})$ be a resolution of identity, i.e., a family of orthogonal projections such that $\sum_{k=0}^{d-1}P_k=Id$. Then we say that $(P_0,\dots,P_{d-1})$ is $\Phi$-cyclic if $\Phi^*(P_k)=P_{k\stackrel{d}{-}1}$, for $k=0,\dots d-1$. The supremum of all $d$ for which there exists a $\Phi$-cyclic resolution of identity $(P_0,\dots,P_{d-1})$ is called the {\bf period} of $\Phi$. If $\Phi$ has period 1 then we call it {\bf aperiodic}.
\end{definition}

With these definitions one may obtain an operator version of the classical Markov chain result \cite{norris}. The following has been presented in \cite{carbone2}, Theorem 4.18.

\begin{theorem}\label{qu_vernorris}
Let $\Phi$ be an irreducible, aperiodic and finite OQRW. For any state $\rho$ the sequence $\Phi^r(\rho)$ converges to the invariant state $\rho_*$, which is unique and faithful.
\end{theorem}

At this point we can state a question. We recall that the classical version of Theorem \ref{qu_vernorris} for column stochastic matrices $P$ \cite{norris} implies that the columns of $P^r$ converge to the unique invariant distribution vector for $P$, as $r\to\infty$. One can ask for an analogous property for quantum channels. By considering the matrix representation $[\Phi]$ of an irreducible, aperiodic, finite quantum channel, does every column converge to a limit distribution? The answer to this question is easily shown to be negative in general. Due to the connection between OQRWs and QTMs, one can ask instead the more refined question: given a QTM $B$ and its matrix expression $B=[B_{j}^i]$, $\sum_i B_{j}^{i*}B_{j}^i=I$, all $j$, does the columns of its iterates converge to some kind of limit vector state? The results in Section \ref{sec5} will produce an affirmative answer, under certain conditions, in terms of a notion of ergodicity of sequences of QTMs.

\medskip

As an example, it is a simple matter to illustrate how classical stochastic matrices may be described in terms of the setting of QTMs. Many aspects of this translation are straightforward, but some details are worth mentioning explicitly.

\begin{example}\label{clbipr}
{\bf (Classical stochastic matrices in the setting of QTMs)}. We say $B$ is a {\bf classical stochastic QTM} if $B_{i}^j=\sqrt{p_{i}^j}I$, where $P=(p_{i}^j)$  is a real stochastic matrix of finite dimension. Then the proof of the following properties are immediate. For simplicity we consider an order 2 QTM $B = (B_i^j)$, $i, j = 1, 2$, where each $B_i^j$ also has order 2, but generalizations to larger (finite) order are clear.

\begin{enumerate}
\item Let $B$ be an order 2 classical stochastic QTM. a) Let $\rho=(\rho_1,\rho_2)^T$ be a vector state and $v=(v_1,v_2)^T$ probability vector such that $v_i=tr(\rho_i)$, $i=1,2$. Then $(Pv)_i=tr(B(\rho))_i$, $i=1,2$. b) If $\rho_1=x_1|1\rangle\langle 1|$ and $\rho_2=x_2|2\rangle\langle 2|$, $x_1+x_2=1$, $x_i>0$, then
\begin{equation}
B(\rho)=\begin{bmatrix} \begin{bmatrix} p_{1}^1x_1 & 0 \\ 0 & p_{2}^1x_2 \end{bmatrix} \\ \\ \begin{bmatrix} p_{1}^2x_1 & 0 \\ 0 & p_{2}^2x_2 \end{bmatrix} \end{bmatrix}
\end{equation}
\item If $P$ is an order 2, aperiodic, irreducible, stochastic real matrix with stationary vector $\pi=(\pi_1,\pi_2)^T$ then a) for the classical stochastic QTM $B$ associated to $P$,
\begin{equation}\label{oi_1}
B^{r}\to C:=\begin{bmatrix} \pi_1I & \pi_1I \\ \pi_2I & \pi_2I\end{bmatrix},\;\;\;r\to\infty
\end{equation}
b) For any vector state $\rho=(\rho_1,\rho_2)^T$, with entries as in
\begin{equation}
\rho_1=\begin{bmatrix} a & b \\ c & d\end{bmatrix},\;\;\;\rho_2=\begin{bmatrix} f & g \\ h & j \end{bmatrix},\;\;\;v=\begin{bmatrix} a+d \\ f+j\end{bmatrix},
\end{equation}
we have for $C$ as in eq. (\ref{oi_1}) that
\begin{equation}\label{typical_express}
C\rho=\begin{bmatrix} \pi_1\begin{bmatrix} a+f & b+g \\ c+h & d+j \end{bmatrix} \\ \\ \pi_2\begin{bmatrix} a+f & b+g \\ c+h & d+j \end{bmatrix} \end{bmatrix}=\begin{bmatrix} \pi_1(\rho_1+\rho_2) \\ \\ \pi_2(\rho_1+\rho_2) \end{bmatrix}
\end{equation}
c) For the vector state $\rho$ with $\rho_1=x_1|1\rangle\langle 1|$ and $\rho_2=x_2|2\rangle\langle 2|$, $x_1+x_2=1$, $x_i>0$, and for the classical stochastic QTM $B$ associated to $P$ which is aperiodic and irreducible,
\begin{equation}\label{adifference}
(B^{r})(\rho)=\begin{bmatrix} \begin{bmatrix} p_{1}^{1(r)}x_1 & 0 \\ 0 & p_{2}^{1(r)}x_2 \end{bmatrix} \\ \\ \begin{bmatrix} p_{1}^{2(r)}x_1 & 0 \\ 0 & p_{2}^{2(r)}x_2 \end{bmatrix} \end{bmatrix}\to \begin{bmatrix} \pi_1\begin{bmatrix} x_1 & 0 \\ 0 & x_2 \end{bmatrix} \\ \\ \pi_2\begin{bmatrix} x_1 & 0 \\ 0 & x_2 \end{bmatrix} \end{bmatrix},\;\;\;r\to\infty
\end{equation}
\end{enumerate}
\end{example}

\qee

\begin{remark}\label{remark_adiff}
The fact that the columns of the iterates of  a QTM cannot converge to a vector state should already be clear from the structure we have: on one hand we must have for a QTM $B=(B_{i}^j)$ that $\sum_j B_{i}^{j*}B_{i}^j=I$ for all $i$, but on the other a vector state $\rho=(\rho_i)$ must satisfy $\sum_i tr(\rho_i)=1$. That is, for a column $(B_{i}^1 \; B_{i}^2 \cdots B_{i}^n)^T$ to be equal to $\rho$ is impossible in general. This is also indicated by expression (\ref{adifference}). Nevertheless, if we consider the case in which all matrices $B_{j}^i$ and $\rho_i$ are one-dimensional then we recover the classical case.
\end{remark}

\section{A spectral technique for QMCs}\label{sec5}

Recall that the {\bf singular values} of an operator $T$ are the square roots of the eigenvalues of the map $T^*T$. In this section we consider a quantum channel $\Phi_B=\Phi$ which is unital, induced by an order $n$ QTM $B$ given by order k matrices, that is, $B=(B_{j}^i)$, $\dim B_{j}^i=k$, all $i,j=1,\dots,n$. We let $\sigma_i(\Phi)$ be the i-th singular value of $\Phi$, $i=1,2,\dots$, arranged in non-increasing order. It is usual to say that $n$ is the number of sites. Note that for any QTM, $\sigma_1(\Phi)=1$ and $\sigma_i(\Phi)\in[0,1]$. Define an inner product for columns,
\begin{equation}
\Big\langle\begin{bmatrix} A_1 \\ A_2 \\ \vdots \\ A_n \end{bmatrix},\begin{bmatrix} C_1 \\ C_2 \\ \vdots \\ C_n \end{bmatrix}\Big\rangle_2:=\langle A_1,C_1\rangle_2+\langle A_2,C_2\rangle_2+\cdots+\langle A_n,C_n\rangle_2,\;\;\;A_i,C_i\in M_k(\mathbb{C})
\end{equation}
where in the right hand side the product appearing is the Hilbert-Schmidt inner product of matrices $\langle A_i,C_i\rangle_2:=tr(A_i^*C_i)$. Denote by $\Vert\cdot\Vert_2$ the norm induced by such product. Note that, as a block matrix, a QTM is an order $kn$ matrix (n rows of order k matrices). Inspired by the asymptotic behavior of the unital examples examined, we would like to compare the distance between the columns of a QTM and the column $\pi=[\frac{1}{n}I \; \cdots\; \frac{1}{n}I]^T$.
\begin{remark}\label{remark_4n2}
Due to the expression for the matrix representation we see that an OQRW $\Phi$ on $n$ sites, with each transition matrix being an order $k$ matrix, is such  that $[\Phi]$ is an order $N_\Phi=(kn)^2$ matrix. To see this, use eq. (\ref{matrep}) with the fact that we can write
\begin{equation}\label{aid_matrep}
\Phi(\rho)=\sum_{i,j} B_{i}^j\rho B_{i}^{j*},\;\;\;B_{i}^j=L_{i}^j\otimes|j\rangle\langle i|,
\end{equation}
for some matrices $L_i^j$ satisfying $\sum_{j} L_{i}^{j*}L_{i}^j=I$, for all $i$. Then, for instance, for a QTM acting on $n$ sites and order 2 transition matrices, the induced OQRW has an order $4n^2$ matrix representation, and a QTM on $2$ sites and order 2 transitions $B=(B_{i}^j)$, $B_{i}^j\in M_2(\mathbb{C})$, $i,j=1,2$, induces an OQRW with an order 16 matrix representation.
\end{remark}
The proof of the following technical result, which is needed later, can be seen in the Appendix.
\begin{pro}\label{saloff_explicado}
Let $B=(B_{i}^j)$ be a unital QTM (i.e., the induced OQRW preserves the identity) and $\Phi_B=\Phi:\bigoplus_{i=1}^n M_k(\mathbb{C})\to \bigoplus_{i=1}^n M_k(\mathbb{C})$ be the induced OQRW on $n$ sites and action given by order $k$ matrices. Let $\sigma_i=\sigma_i(\Phi)$, $i=1,\dots,n$ be the singular values of $\Phi$. Let $\{\eta_i\}_{i=1}^{N_\Phi}$ denote an orthonormal basis of eigenstates for $\Phi^*\Phi$, associated to eigenvalues arranged in non-increasing order, with $\eta_1$ given by
\begin{equation}
\eta_1=\frac{1}{\sqrt{kn}}\Big(I\otimes |1\rangle\langle 1|+\cdots +I\otimes |n\rangle\langle n|\Big), \;\;\;I=I_k\in M_k(\mathbb{C}),
\end{equation}
in the canonical basis. Then for all $j=1,2,\dots,n$,
\begin{equation}\label{saloff_b1}
\sum_{i=1}^n \Big\Vert B_{j}^{i*}B_{j}^i-\frac{I}{n}\Big\Vert_2^2=\sum_{i=2}^{N_\Phi} |d_{ij}|^2\sigma_i^2,
\end{equation}
where for each $j$, $\begin{bmatrix} 0 & \cdots & I & \cdots & 0\end{bmatrix}^T=\sum_i d_{ij}\eta_i$, with $I$ the order $k$ identity appearing in the $j$-th position.
\end{pro}

Now let $\mathcal{B}=\{B_i\}_{i\in\mathbb{N}}$ denote a sequence of QTMs (i.e., a QMC). We use the notation $\mathcal{B}_{i,i}=I$ and
\begin{equation}
\mathcal{B}_{p,q}=B_{p+1}\cdots B_q,\;\;\;p\leq q,
\end{equation}
where by the above product we mean the QTM associated to the product of matrix representations of the QTMs $B_{p+1},\cdots, B_q$ (equivalently, the product of the associated OQRWs). In the special case of a homogeneous QMC $\mathcal{B}=\{B^r\}_{r=1}^\infty$ the expression $\mathcal{B}_{p,q}$ equals $\mathcal{B}_{0,q-p}=B^{q-p}$ (product of matrix representations). An advantage on employing matrix representations is that this allows us to make use of well-known results on the singular values of the product of matrices. We make this choice even though this is a matrix which is larger than, say, the Kraus matrices of a channel. Moreover, we denote by $B_i(l,m)$ the matrix appearing in the $(l,m)$-th position of $B_i$ (and not the $(l,m)$-th numeric entry of the QTM).

\begin{definition}\label{ergodef}
Fix a vector state $\rho_\pi=[\rho_1 \dots \rho_n]^T$, $\sum_i tr(\rho_i)=1$, with $\rho_i\in M_k(\mathbb{C})$. Let $\mathcal{Q}=\{Q_1,\dots, Q_q\}$ be a finite family of QTMs, all admitting $\rho_\pi$ as an invariant measure. We say that the pair $(\mathcal{Q},\rho_\pi)$ is {\bf ergodic} if for any QMC defined by the sequence of QTMs $\mathcal{B}=\{B_i\}_{i\in\mathbb{N}}$  with invariant measure $\rho_\pi$, such that $B_i\in\mathcal{Q}$ for infinitely many $i$'s, we have for all $i,j,k\in\{1,\dots,n\}$,
\begin{equation}\label{linesequal}
\lim_{r\to\infty} \mathcal{B}_{0,r}(i,j)-\mathcal{B}_{0,r}(i,k)=0.
\end{equation}
\end{definition}
That is, we verify whether all columns of the resulting product are becoming equal. Now we define a notion of distance between columns. In the case of unital channels we will fix the maximally mixed column $\pi=[I/n \cdots I/n]^T$, $I$ being the identity matrix of order $k$, and calculate the distance from a given column to $\pi$. It is worth noting once again that $\pi$, being a column of a QTM, is not a vector state. However, we will see that this produces a consistent limit theorem (see eq. (\ref{oi_1}) and Remark \ref{remark_adiff}). Define
\begin{equation}\label{goodmetric1}
d_2(\mu;\pi)=\Big(\sum_{i=1}^n\Big\Vert \mu(i)\mu(i)^*-\frac{I}{n}\Big\Vert_2^2\Big)^{1/2},
\end{equation}

\medskip

Now we recall the following fact. By exercise 4, p. 182 \cite{hj2}, we have that for any set of matrices $A_1,\dots,A_m\in M_k(\mathbb{C})$, $m\geq 2$, $l=1,\dots,k$,
$\sum_{i=1}^l\sigma_i(A_1\cdots A_m)\leq \sum_{i=1}^l\sigma_i(A_1)\cdots\sigma_i(A_m)$. In particular, for any set of matrices, and noting that $\sigma_1(A_1\cdots A_m)=1$,
\begin{equation}\label{basichorn1}
\sigma_2(A_1\cdots A_m)\leq\Pi_{i=1}^m\sigma_2(A_i)
\end{equation}

The following is the QTM version of a technical result proved in \cite{saloff}, and will be needed for Theorem \ref{bigbigt}.
\begin{pro}
Let $\mathcal{B}=(B_i)_1^\infty$ be a sequence of unital QTMs on $n$ sites with matrices on $M_k(\mathbb{C})$. For each $j$ let $\sigma_i(B_j)$, $i=1,\dots,N_{\Phi_B}$ be the singular values of the OQRW induced by $B_j$. Then for every $j\geq 1$, and every $m\geq 1$, there is a constant $C(j,n)$ such that
\begin{equation}\label{saloff_2}
d_2(\mathcal{B}_{0,m}(\cdot,j);\rho_\pi)\leq C(j,n)\prod_{l=1}^m\sigma_2(B_l)
\end{equation}
\end{pro}
{\bf Proof.} We apply (\ref{saloff_b1}) with $\Phi=\mathcal{B}_{0,m}$. By definition, $\sigma_j(\mathcal{B}_{0,m})\leq\sigma_2(\mathcal{B}_{0,m})$, $j=2,\dots ,n$ so we get
\begin{equation}
d_2(\mathcal{B}_{0,m}(\cdot,j);\rho_\pi)=\Big(\sum_{i=2}^{N_{\Phi_B}}|d_{ij}|^2\sigma_i^2\Big)^{1/2}\leq\sigma_2(B_1\cdots B_m)\Big(\sum_{i=2}^{N_{\Phi_B}} |d_{ij}|^2\Big)^{1/2}\leq C(j,n)\prod_{l=1}^m\sigma_2(B_l),
\end{equation}
where $C(j,n)=\sum_{i=2}^{N_{\Phi_B}} |d_{ij}|^2$. Note that the $d_{ij}$ depend on $m$ as well, but $C(j,n)$ does not increase arbitrarily with $m$, due to the bound $|d_{ij}|\leq 1$, for all $m$.


\qed

Now recall that the {\bf geometric multiplicity} $\gamma_\Phi(\lambda)$ of an eigenvalue $\lambda$ is the dimension of the eigenspace associated with $\lambda$, i.e., $\dim\ker(\Phi-\lambda I)$, the maximum number of vectors in any linearly independent set of eigenvectors with that eigenvalue. The {\bf algebraic multiplicity} $\mu_\Phi(\lambda)$ of $\lambda$ is its multiplicity as a root of the characteristic polynomial and it is well-known that $\gamma_\Phi(\lambda)\leq \mu_\Phi(\lambda)$. We recall the following important basic fact, the proof of which is described by M. Wolf \cite{wolf}.
\begin{pro}\label{wolf1}
(Trivial Jordan blocks for peripheral spectrum). Let $\Phi$ be a trace-preserving (or unital) positive linear map. If $\lambda$ is an eigenvalue of $\Phi$ with $|\lambda|=1$ then its geometric multiplicity equals its algebraic multiplicity, i.e., all Jordan blocks for $\lambda$ are one-dimensional.
\end{pro}

\begin{remark}\label{rafterwolf}
As a complement to our discussion, by \cite{carbone}, Proposition 3.5, if we have an irreducible quantum channel $\Phi$ then for every eigenvalue $\lambda$ with $|\lambda|=1$ we have $\dim \ker(\Phi-\lambda I)=1$. By Proposition \ref{wolf1}, for an irreducible quantum channel we have that 1 is an eigenvalue of algebraic multiplicity 1.
\end{remark}

By considering the Hilbert-Schmidt inner product on $B(\mathcal{H})$, the adjoint of a unital quantum channel $\Phi(\rho)=\sum_i V_i\rho V_i^*$ is the unital channel $\Phi^*(\rho)=\sum_i V_i^*\rho V_i$. The {\bf square modulus} of a unital channel $\Phi$ is the unital channel $\Phi\Phi^*$. It is clear that the square modulus is a self-adjoint non-negative operator, that is, $\langle A,\Phi\Phi^*(A)\rangle=\langle\Phi^*(A),\Phi^*(A)\rangle\geq 0$, $\forall\;A\in B(\mathbb{H})$. This implies that $\Phi^*\Phi$ can be diagonalized and has only non-negative eigenvalues \cite{burgarth}.

\medskip

The following is the QMC version of the theorem presented in \cite{saloff}. As mentioned in the Introduction this result is somewhat expected and the method of proof (seen in the Appendix) may be of independent interest and follows with little difficulty.

\begin{theorem}\label{bigbigt}
Let $\mathcal{Q}=\{Q_1,\dots, Q_q\}$ be a finite family of unital QTMs. Then the pair $(\mathcal{Q},\rho_\pi)$, $\rho_\pi$ being the maximally mixed vector state, is ergodic in the sense of Definition \ref{ergodef} if and only if $\sigma_2(Q_j)<1$ for each $j\in\{1,\dots,q\}$.
\end{theorem}

\begin{example}\label{relev_ex1}
Let
\begin{equation}
V_1=\frac{1}{\sqrt{3}}\begin{bmatrix} 0 & 1 \\ 1 & 0 \end{bmatrix},\;\;\;V_2=\frac{1}{\sqrt{3}}\begin{bmatrix} 0 & -i \\ i & 0 \end{bmatrix},\;\;\;V_3=\frac{1}{\sqrt{3}}\begin{bmatrix} 1 & 0 \\ 0 & -1 \end{bmatrix}
\end{equation}
Then $\sum_i V_i^*V_i=I$ and we can build an order 3 QTM with the above matrices in the following way. Let
\begin{equation}
B=\begin{bmatrix} V_1 & V_2 & V_3 \\ V_2 & V_3 & V_1 \\ V_3 & V_1 & V_2 \end{bmatrix}.
\end{equation}
In terms of OQRWs this corresponds to define $\rho\mapsto\sum_{i,j} M_{ij}\rho M_{ij}^*$, $M_{1i}=V_i\otimes E_{1i}$, $i=1,2,3$, where $(E_{ij})_{kl}=\delta_{(i,j),(k,l)}$ are the order 3 matrix units, and analogously for the other rows. Then a calculation shows that $\sigma_1(B)=1$ and $\sigma_2(B)=\frac{2}{3}<1$. We conclude that the pair $(\{B\},\rho_\pi)$, $\rho_\pi$ being the maximally mixed state, is ergodic by Theorem \ref{bigbigt}. We note that in this example one can calculate the singular values of its representation matrix, which has order $(nk)^2=(3\cdot 2)^2=36$, see Remark \ref{remark_4n2}. The singular values $1, 2/3, 1/3$ and $0$ have multiplicities $1, 6, 3$ and $26$, respectively. Noting the fact that the matrix representation is quite large already for the case of 3 sites, one may ask for smaller matrices which represent the same OQRW. This is indeed possible and a detailed study on this matter will be discussed in a future work.

\end{example}

\qee

\begin{example}\label{relev_ex2}
If a classical bistochastic QTM of order 2 $B=(B_{ij})$, $B_{ij}\in M_2(\mathbb{C})$ belongs to a finite set $\mathcal{Q}$ then no pair $(\mathcal{Q},\rho_\pi)$ is ergodic, since $\sigma_2(B)=1$ as a calculation shows. For a different example, let
\begin{equation}
V_1=\frac{1}{\sqrt{3}}\begin{bmatrix} 1 & 1 \\ 0 & 1 \end{bmatrix},\;\;\; V_2=\frac{1}{\sqrt{3}}\begin{bmatrix} 1 & 0 \\ -1 & 1 \end{bmatrix}
\end{equation}
Then $\sum_i V_i^*V_i=I$ and we can build an order 2 QTM with the above matrices by writing
\begin{equation}
B=\begin{bmatrix} V_1 & V_2 \\ V_2 & V_1 \end{bmatrix}.
\end{equation}
We are able to write the corresponding OQRW as in the above example. Then a calculation shows that $\sigma_1(B)=\sigma_2(B)=1$. We conclude that if $B\in\mathcal{Q}$ then no pair $(\mathcal{Q},\rho_\pi)$, will be ergodic ($\rho_\pi$ being the maximally mixed state). As another observation,  consider the 1-qubit quantum channel $\Phi(\rho)=V_1\rho V_1^*+V_2\rho V_2^*$ which has 1 as the the unique eigenvalue of modulus 1, and has multiplicity 1. This shows that one can find a quantum channel that has a unique fixed point, which is attractive (a mixing channel, in the terminology of \cite{burgarth}), but such that the induced QTM is not ergodic in our sense.
\end{example}

\qee

\subsection{Ergodicity on noncommutative $L^1$-spaces}

As a complementary discussion on ergodicity we would like to comment on QTMs in terms of the setting studied by F. Mukhamedov \cite{mukhamedov}, that is, in a noncommutative $L^1$-space setting. In that context the mentioned author is able to define a Dobrushin coefficient and then analyze ergodicity properties of certain operators. We will discuss some of these results in light of our QTM description and we will see that many facts are easily examined on finite-dimensional Hilbert spaces when proper adaptations are made.

\medskip

We will consider the algebra $B(\mathcal{H}\otimes\mathcal{K})$ of bounded linear operators acting on $\mathcal{H}\otimes\mathcal{K}$, where $\mathcal{H}$ and $\mathcal{K}$ finite dimensional (see Section \ref{bigrem}). As mentioned before, the vector states described in previous sections form a convex subset of $B(\mathcal{H}\otimes\mathcal{K})$. In finite dimensions all norms are equivalent with, for instance, $\Vert \cdot\Vert_2\leq \Vert \cdot\Vert_1$, see  \cite{watrous}, and $\Vert \cdot\Vert_2\leq \sqrt{n}\Vert \cdot\Vert_\infty$, $\Vert \cdot\Vert_\infty$ being the usual maximum norm on matrices \cite{benatti,hj1}.  


\medskip

As discussed in Section \ref{sec2}, every vector state $\rho=(\rho_i)$, $\sum_{i=1}^n tr(\rho_i)=1$, $\rho_i\in M_k(\mathbb{C})$ can be identified with $\rho=\sum_i \rho_i\otimes |i\rangle\langle i|$, and since $\omega(\rho):=\sum_i tr(\rho_i)=tr(\sum_i \rho_i\otimes |i\rangle\langle i|)$, we have that $\hat{\omega}=\frac{1}{kn}\omega$ is a faithful state functional on the vector states for a QTM. For a given vector state $\rho$, define $T_\rho$ acting on vector states as follows,
\begin{equation}
T_\rho(X):=tr(X)\rho, \;\;\;X=\sum_{i=1}^n X_i\otimes |i\rangle\langle i|,\;\;\;X_i\in M_k(\mathbb{C})
\end{equation}
Following \cite{mukhamedov}, a QTM $A$ is {\bf uniformly ergodic} if there exists an element $Y$ such that
\begin{equation}\label{unif_erg}
\lim_{r\to\infty}\Vert \mathcal{A}_{m,r}-T_Y\Vert_\infty=0,
\end{equation}
for all $m\geq 0$.
Note that $Y$ plays the role of an  equilibrium state.
Also following \cite{mukhamedov}, a QTM $A$ is {\bf weakly ergodic} if for every $k\in\mathbb{N}\cup\{0\}$ we have
\begin{equation}\label{weak_ergd}
\lim_{r\to\infty}\sup_{\rho,\eta\in \mathcal{D}(\mathcal{H})}\Vert \mathcal{A}_{k,r}\rho-\mathcal{A}_{k,r}\eta\Vert_1=0,
\end{equation}
where $\Vert X\Vert_1=\hat{\omega}(|X|)$ and $|X|=\sqrt{X^*X}$.

\medskip

Now we state a result that is also described by \cite{mukhamedov}, with notation adapted for our purpose, considering  homogeneous QMCs.
Instead of considering the norm induced by the Hilbert-Schmidt norm, as it has been made previously, we will consider $L^1$-spaces, and state results in terms of the norm $\Vert \cdot\Vert_1$. One of the facts that can be discussed at this point is a relation between weak ergodicity, uniform ergodicity and singular values of a QTM. This gives us an asymptotic notion which is closely related to the one presented in this work. In order to do that, we recall some definitions. Let
\begin{equation}
\mathcal{Z}:=\{X=\sum_{i=1}^n X_i\otimes |i\rangle\langle i|: tr(X)=0\}
\end{equation}
and for $T$ a linear operator define
\begin{equation}\label{dobrush_coeff}
\delta(T):=\sup_{X\in \mathcal{Z}, X\neq 0} \frac{\Vert TX\Vert_1}{\Vert X\Vert_1}
\end{equation}
the {\bf Dobrushin ergodicity coefficient} of $T$. Basic properties of this coefficient can be seen in \cite{mukhamedov}, where it is proven that
\begin{equation}\label{equivdobrush_coeff}
\delta(T)=\sup_{\rho,\eta\in \mathcal{D}(\mathcal{H})}\frac{\Vert T\rho-T\eta\Vert_1}{2}.
\end{equation}

\begin{theorem}\label{theomukh}
\cite{mukhamedov} Let $B$ be a finite dimensional QTM. The following are equivalent: a) The homogeneous QMC $\{B^r\}_{r=1}^\infty$ is weakly ergodic. b) There exists $s\in[0,1)$ and $n_0\in\mathbb{N}$ such that $\delta(B^{n_0})\leq s$. c) $\mathcal{B}=\{B^r\}_{r=1}^\infty$ is uniformly ergodic.
\end{theorem}

 In the homogeneous case,  Definition \ref{ergodef} is reduced to

\begin{definition}\label{ergodef-h}
Fix a vector state $\rho_\pi=[\rho_1 \dots \rho_n]^T$, $\sum_i tr(\rho_i)=1$, with $\rho_i\in M_k(\mathbb{C})$. Let $B$ be a QTM, that admits $\rho_\pi$ as an invariant measure. We say that the pair $(B,\rho_\pi)$ is {\bf ergodic} if for all $i,j,k\in\{1,\dots,n\}$,
\begin{equation}\label{linesequal-h}
\lim_{r\to\infty} B^r(i,j)-B^r(i,k)=0.
\end{equation}
\end{definition}

The following corollary is immediate from Theorem \ref{theomukh} and Definition \ref{ergodef-h}:

\begin{cor}\label{bogcor}
Suppose one of the conditions of Theorem \ref{theomukh} holds for a given homogeneous QMC $\mathcal{B}=\{B^r\}_{r=1}^\infty$. Then $\mathcal{B}$ is ergodic in the sense of Definition \ref{ergodef-h}. As a consequence, $\sigma_2(B)<1$.
\end{cor}

On the other hand, assuming that a QTM $B$ is ergodic in the sense of Definition \ref{ergodef} then it is a simple matter to show that this does not imply weak ergodicity.  In fact, let $B$ be the order 2 maximally mixed QTM and note that for every $\rho$,
\begin{equation}
B(\rho)=\begin{bmatrix} \frac{I}{\sqrt{2}} & \frac{I}{\sqrt{2}} \\ \frac{I}{\sqrt{2}} & \frac{I}{\sqrt{2}}\end{bmatrix}\begin{bmatrix} \rho_1 \\ \rho_2\end{bmatrix}=\begin{bmatrix} \frac{1}{2}(\rho_1+\rho_2) \\ \frac{1}{2}(\rho_1+\rho_2)\end{bmatrix}.
\end{equation}
Let
\begin{equation}
\rho=\begin{bmatrix} \rho_1 & \rho_2 \end{bmatrix}^T=\begin{bmatrix} I/4 & I/4 \end{bmatrix}^T,\;\;\;\eta=\begin{bmatrix} \eta_1 & \eta_2 \end{bmatrix}^T=\begin{bmatrix} \begin{bmatrix} 1 & 0 \\ 0 & 0 \end{bmatrix} & \begin{bmatrix} 0 & 0 \\ 0 & 0 \end{bmatrix} \end{bmatrix}^T
\end{equation}
Then
\begin{equation}
B\rho=\rho,\;\;\; B\eta=\frac{1}{2}\begin{bmatrix} \begin{bmatrix} 1 & 0 \\ 0 & 0 \end{bmatrix} &\begin{bmatrix} 1 & 0 \\ 0 & 0 \end{bmatrix}\end{bmatrix}^T\;\;\Longrightarrow \;\;B\rho-B\eta=\begin{bmatrix} -\frac{1}{4} & 0 \\ 0 & \frac{1}{4}\end{bmatrix}
\end{equation}
Hence, we are able to perform a calculation of (\ref{weak_ergd}) which produces a positive number. This shows that the homogeneous QMC induced by $B$ is ergodic in the sense of Definition \ref{ergodef}, but is not weakly ergodic.

\section{Quantum hitting time}\label{zxsec3}

In this section we begin our discussion on hitting times for OQRWs. As previously discussed the description can be made in terms of QTMs as well, but we will fix notation for quantum channels and its associated quantum trajectories. Unlike QTMs, we usually assume an infinite number of sites for the OQRWs considered (walks on $\mathbb{Z}$ being a natural example) unless otherwise stated.

\medskip

First we make a remark on general quantum channels. Let
\begin{equation}
\Lambda(\rho)=\sum_k\langle e_k|U[\rho\otimes|e_0\rangle\langle e_0|]U^*|e_k\rangle=\sum_i V_k\rho V_k^*,\;\;\;V_k:=\langle e_k|U|e_0\rangle,
\end{equation}
where $U$ is unitary. Suppose a measurement of the environment is performed in the basis $|e_k\rangle$ after $U$ is applied. By the principle of implicit measurement, such measurement affects only the state of the environment and does not change the state of the principal system. If $\rho_k$ is the state of the principal system, given that outcome $k$ has occurred, then $\rho_k$ is proportional to $V_k\rho V_k^*$. Normalizing,
\begin{equation}
\rho_k=\frac{V_k\rho V_k^*}{tr(V_k\rho V_k^*)}
\end{equation}
and the probability of outcome $k$ is given by $p(k)=tr(|e_k\rangle\langle e_k|U(\rho\otimes|e_0\rangle\langle e_0|)U^*|e_k\rangle\langle e_k|)=tr(V_k\rho V_k^*)$. Therefore,
\begin{equation}\label{motiv1}
\Lambda(\rho)=\sum_k V_k\rho V_k^*=\sum_k tr(V_k\rho V_k^*)\frac{V_k\rho V_k^*}{tr(V_k\rho V_k^*)}=\sum_k p(k)\rho_k
\end{equation}
This gives us the following physical interpretation of the channel: the action of the quantum operation is equivalent to taking the state $\rho$ and randomly replacing it by $V_k\rho V_k^*/tr(V_k\rho V_k^*)$ with probability $tr(V_k\rho V_k^*)$. This kind of calculation has been described in \cite{nielsen} and has also been explored in the context of quantum iterated function systems \cite{bllt2,lozinski}. A point to be examined in this work is the study of functionals of matrices: given (\ref{motiv1}), one may apply a linear functional $f$ on both sides to obtain
$f(\Phi(\rho))=\sum_k p(k)f(\rho_k)$ and a familiar choice is to take $f_V(\rho)=tr(V\rho V^*)$. We will use this together with other notions in order to study hitting times, which is to be presented next. For concrete calculations, we will consider walks on $\mathbb{Z}$ and on certain situations make use of path counting techniques.

\medskip

Before discussing hitting times, we recall that F. Gr\"unbaum et al. have presented a definition of quantum recurrence for discrete time unitary evolutions \cite{werner}. In \cite{cfrr} this notion has been adapted to the setting of OQRWs and below we briefly review this construction.

\medskip

Let $\Phi$ be an OQRW, fix $\rho^{(0)}\otimes|0\rangle\langle 0|$ as the initial state and let $Q$ be an operator acting on states of the random walk in the following way: if $\rho^{(k)}=\Phi^k(\rho^{(0)})$ is a state then $Q\rho^{(k)}$ equals $(I-|0\rangle\langle 0|)\rho^{(k)}$. That is, the paths returning to zero are removed at every iteration via a projection to its complement. For instance, let $\Phi$ denote the nearest neighbor OQRW on $\mathbb{Z}$ induced by matrices $L$ and $R$ such that $L^*L+R^*R=I$, i.e.,
\begin{equation}
\Phi(\rho)=\sum_{i\in\mathbb{Z}}\Big(L\rho_{i+1}L^*+R\rho_{i-1}R^*\Big)\otimes|i\rangle\langle i|,\;\;\;\rho=\sum_{i\in\mathbb{Z}} \rho_i\otimes |i\rangle\langle i|
\end{equation}
Define
\begin{equation}\label{secondst}
\alpha=L^2\rho_0L^{2*}\otimes|-2\rangle\langle -2|+(LR\rho_0R^*L^*+RL\rho_0L^*R^*)\otimes |0\rangle\langle 0|+R^2\rho_0 R^{2*}\otimes|2\rangle\langle 2|
\end{equation}
Then the probability of occurrence of state $|-2\rangle$, $|0\rangle$ and $|2\rangle$ are, respectively, $tr(L^2\rho_0L^{2*})$, $tr(LR\rho_0R^*L^*+RL\rho_0L^*R^*)$, $tr(R^2\rho_0 R^{2*})$ and we get
\begin{equation}
Q\alpha=L^2\rho_0L^{2*}\otimes|-2\rangle\langle -2|+R^2\rho_0 R^{2*}\otimes|2\rangle\langle 2|\;.
\end{equation}
Now for a given state $\rho$, let $S_n(\rho)$ denote the probability of occurrence of a site other than zero at time $n$,
\begin{equation}
S_n(\rho):=\mu((Q\Phi)^n(\rho)),
\end{equation}
where $\mu(\rho)=\mu(\sum_i \rho_i\otimes |i\rangle\langle i|):=\sum_i tr(\rho_i)$. Define the {\bf return probability} by
$R:=1-\lim_{n\to\infty} S_n(\rho)$. We say that site $|0\rangle$ is {\bf recurrent} for $\Phi$ if $R=1$ for all $\rho=\rho_0\otimes |0\rangle\langle 0|$, and we say it is {\bf transient} otherwise. We give the analogous definition for walks starting at sites other than $|0\rangle$.
As remarked in \cite{cfrr}, when considering a quantum system together with the probabilistic notion of recurrence, if one has to check whether a system has reached a certain state then such inspection is a measurement which modifies the system. Our approach here is the same as the one taken in \cite{werner}, which is to consider a notion of recurrence that {\bf includes} the system monitoring into the description. We also note that the definition above resembles a kind of {\bf site} recurrence, that is, recurrence of a particular position on $\mathbb{Z}$. One should compare this notion with {\bf state} recurrence \cite{bourg}, and it is worth mentioning that there exist many other approaches to quantum recurrence, as this topic can be discussed in terms of operator algebras \cite{accardi1,accardi2}, Markov semigroups \cite{fagnola} and spectral theory \cite{oliveira}, also see the references in \cite{werner}.

\medskip

Inspired by the recurrence definition given above, it is a simple matter to define a notion of hitting time in a similar way, that is, including the system monitoring into the description. As it happens in \cite{cfrr}, many concrete calculations will rely on some path counting technique.

\begin{definition} Let $\Phi$ be an OQRW. The {\bf probability of first visit to site $j$ at time $r$}, starting at $\rho_i\otimes|i\rangle\langle i|$ is denoted by $b_r(\rho_i;j)$. This is the sum of the traces of all paths allowed by $\Phi$ starting at $\rho_i\otimes |i\rangle\langle i|$ and reaching $j$ for the first time at the $r$-th step (an analytic expression is given below). The probability starting from $\rho_i\otimes |i\rangle\langle i|$ that the walk ever hits site $j$ is
\begin{equation}
h_{i}^j(\rho_i)=\sum_{r=1}^\infty b_r(\rho_i;j),\;\;\;i\neq j
\end{equation}
and $h_{i}^i(\rho_i)=1$. This is the {\bf probability of visiting site $j$}, given that the walk started at site $i$.
\end{definition}

The time of first visit will then be used as an synonym for {\bf hitting time}. Also, it makes sense to consider the probability of first visit to a set $A$ and denote it by $h_i^A(\rho)$, with $h_i^A(\rho)=1$ if $i\in A$.

\medskip

Let $\pi_r(j;A)$ be the set of all products of $r$ matrices corresponding to a sequence of sites that a walk is allowed to perform with $\Phi$, beginning at site $|j\rangle$, first reaching set $A$ in the $r$-th step. We remark that $\pi_r(j,A)\cap \pi_s(j,A)=\emptyset$ if $r\neq s$. For instance, for the OQRW (\ref{oqrwdefinition}) acting on $\mathbb{Z}$ and, reading matrices from right to left, we have $B_{3}^4B_{2}^3B_{1}^2B_{2}^1B_{1}^2\in\pi_5(1;\{4\})$, as this corresponds to moving right, left and then right 3 times. Let $\pi(j;A)=\cup_{r=1}^\infty \pi_r(j;A)$ and note that if $i\notin A$ then
\begin{equation}\label{umadef_imp}
h_i^A(\rho_i)=\sum_{r=1}^\infty b_r(\rho_i;A)=\sum_{r=1}^\infty\sum_{C\in\pi_r(j;A)}tr(C\rho_i C^*)=\sum_{C\in\pi(j;A)}tr(C\rho_i C^*),
\end{equation}
and $h_i^A(\rho_i)=1$ if $i\in A$.

\begin{definition} For fixed initial state and final site, the {\bf expected hitting time} is
\begin{equation}
k_{i}^j(\rho_i)=\sum_{r=1}^\infty rb_r(\rho_i;j).
\end{equation}
\end{definition}

\section{Minimal solution for hitting time problems}\label{zxsec4}

As a motivation we review a classical example \cite{norris}.

\begin{example}\label{walk4sitecla}
Consider a classical random walk on 4 sites, with transition probabilities given by the following. Write $P=(p_{ij})$ with $p_{ij}$ denoting the probability of reaching site $j$ in one step, given that it was in $i$:
\begin{equation}\label{clmatrix}
P=\begin{bmatrix} 1 & 0 & 0 & 0 \\ \frac{1}{2} & 0 & \frac{1}{2} & 0 \\ 0 & \frac{1}{2} & 0 & \frac{1}{2} \\ 0 & 0 & 0 & 1 \end{bmatrix}
\end{equation}
That is, states $1$ and $4$ are absorbing. Let $h_i^4$ denote the probability of ever reaching site $4$, given that the walk started at site $i$.
Then, using the law of total probability (i.e., conditioning on the first step) we have $h_2^4=\frac{1}{2}h_1^4+\frac{1}{2}h_3^4$, $h_3^4=\frac{1}{2}h_2^4+\frac{1}{2}h_4^4$ and a simple calculation shows that $h_2^4=\frac{1}{3}$: starting from 2, the probability of absorption in $4$ is $1/3$. This kind of calculation and similar ones can be made via well-known recurrence relation methods.
\end{example}
\qee

A natural question is: can we obtain a quantum generalization of the above example? Let us make a calculation. We have
$$\sum_j h_j^A\Big(\frac{B_i^j\rho B_i^{j*}}{tr(B_i^j\rho B_i^{j*})}\Big)tr(B_i^j\rho B_i^{j*})=\sum_j\sum_{C\in\pi(j;A)} tr\Big(\frac{CB_i^j\rho B_i^{j*}C^*}{tr(B_i^j\rho B_i^{j*})}\Big)tr(B_i^j\rho B_i^{j*})$$
\begin{equation}\label{cfeq1}
=\sum_j\sum_{C\in\pi(j;A)} tr(CB_i^j\rho B_i^{j*}C^*)=\sum_{D\in\pi(i;A)}tr(D\rho D^*)=h_i^A(\rho)
\end{equation}
Note that a classical expression is recovered \cite{norris} when we take order 1 density matrices thus eliminating the matrix dependence of $h_i^A$ for any given site $i$: in this particular case we have $h_i^A(\rho_i)=h_i^A$, for $B_i^j=\sqrt{p_{ij}}I$ we get $tr(B_i^j\rho_i B_i^{j*})=p_{ij}$, and (\ref{cfeq1}) becomes
\begin{equation}\label{eqnor5}
\sum_j p_{ij}h_j^A=h_i^A,\;\;\; i\notin A
\end{equation}
and $h_i^A=1$ if $i\in A$. Note that eq. (\ref{eqnor5}) is just the matrix equation $Ph=h$. As a conclusion we have that for any density $\rho$ the vector of hitting times $(h_i^A)_{i\in\mathbb{Z}}$, for any set $A$ is a solution to the functional equation
\begin{equation}\label{funct_eq}
\sum_j tr(B_i^j\rho B_i^{j*})x_j\Big(\frac{B_i^j\rho B_i^{j*}}{tr(B_i^j\rho B_i^{j*})}\Big)=x_i(\rho),\;\;\;i\notin A
\end{equation}
and $x_i(\rho)=1$ if $i\in A$ for every $\rho$. Motivated by (\ref{eqnor5}), we call (\ref{funct_eq}) the {\bf open quantum version} of $Px=x$.

\begin{remark}
We note that in equation (\ref{funct_eq}), under the assumption that $x_j$ is linear, we have an evident cancellation of terms. One of the reasons we have chosen not to perform the simplification is to emphasize the statistical interpretation of certain functionals, as discussed with eq. (\ref{motiv1}): a certain choice occurs, and this is done with a certain probability. Nevertheless, in the rest of this work we will occasionally cancel terms in certain expressions of this kind whenever it is convenient.
\end{remark}

Now we will prove the main result of this section.
\begin{theorem}\label{minimality_th}
For every density $\rho\otimes |i\rangle\langle i|$ the vector of hitting probabilities $(h_i^A(\rho))_{i\in\mathbb{Z}}$ is the minimal nonnegative solution to the system
\begin{equation}
\left\{
\begin{array}{ll}
x_i(\rho)=1 & \textrm{ if } i\in A  \\
\sum_j tr(B_i^j\rho B_i^{j*})x_j\Big(\frac{B_i^j\rho B_i^{j*}}{tr(B_i^j\rho B_i^{j*})}\Big)=x_i(\rho)
 & \textrm{ if } i\notin A
\end{array} \right.
\end{equation}
\end{theorem}
Minimality means that if $x=(x_i)_{i\in\mathbb{Z}}$, is another solution with $x_i(\rho)\geq 0$ for all $i$ then $x_i(\rho)\geq h_i^A(\rho)$ for all $i$. Solutions are assumed to be linear functionals acting on the space of density matrices $D(\mathcal{H}\otimes\mathcal{K})$ associated to the given OQRW $\Phi$.

\medskip

{\bf Proof.} Suppose $(x_i)_{i\in\mathbb{Z}}$ is any solution to (\ref{funct_eq}). Then $x_i=h_i^A=1$ if $i\in A$. Suppose $i\notin A$, then for any density $\rho$ we have
\begin{equation}
x_i(\rho)=\sum_j tr(B_i^j\rho B_i^{j*})x_j\Big(\frac{B_i^j\rho B_i^{j*}}{tr(B_i^j\rho B_i^{j*})}\Big)=\sum_{j\in A}tr(B_i^j\rho B_i^{j*})+\sum_{j\notin A}tr(B_i^j\rho B_i^{j*})x_j\Big(\frac{B_i^j\rho B_i^{j*}}{tr(B_i^j\rho B_i^{j*})}\Big)
\end{equation}
Substitute for $x_j$ to obtain
\begin{equation}
x_i(\rho)=\sum_{j\in A}tr(B_i^j\rho B_i^{j*})\;+
\end{equation}
$$+\sum_{j\notin A}tr(B_i^j\rho B_i^{j*})\Bigg(\sum_{k\in A}tr(B_j^k\frac{B_i^j\rho B_i^{j*}}{tr(B_i^j\rho B_i^{j*})} B_j^{k*})+\sum_{k\notin A}tr(B_j^k\frac{B_i^j\rho B_i^{j*}}{tr(B_i^j\rho B_i^{j*})} B_j^{k*})x_k\Big(\frac{B_j^k\frac{B_i^j\rho B_i^{j*}}{tr(B_i^j\rho B_i^{j*})} B_j^{k*}}{tr(B_j^k\frac{B_i^j\rho B_i^{j*}}{tr(B_i^j\rho B_i^{j*})} B_j^{k*})}\Big) \Bigg)$$
In terms of quantum trajectories $(\rho_n,X_n)$, we may write the above expression as
$$x_i(\rho)=P(X_1\in A)+P(X_1\notin A,X_2\in A)+\sum_{j,k\notin A}tr(B_i^j\rho B_i^{j*})tr(B_j^k\frac{B_i^j\rho B_i^{j*}}{tr(B_i^j\rho B_i^{j*})} B_j^{k*})x_k\Big(\frac{B_j^kB_i^j\rho B_i^{j*} B_j^{k*}}{tr(B_j^kB_i^j\rho B_i^{j*} B_j^{k*})}\Big)$$
\begin{equation}
=P(X_1\in A)+P(X_1\notin A,X_2\in A)+\sum_{j,k\notin A}tr(B_j^kB_i^j\rho B_i^{j*}B_j^{k*})x_k\Big(\frac{B_j^kB_i^j\rho B_i^{j*} B_j^{k*}}{tr(B_j^kB_i^j\rho B_i^{j*} B_j^{k*})}\Big)
\end{equation}
We remark that we should actually write $P(X_1\in A)=P_{\rho\otimes |i\rangle\langle i|}(X_1\in A)$, or a similar notation, if one wants to make explicit the initial density matrix. We will omit the subscripts for simplicity.
By induction,
$$x_i(\rho)=P(X_1\in A)+P(X_1\notin A,X_2\in A)+\cdots+P(X_1\notin A,\dots,X_{n-1}\notin A,X_n\in A)\;+$$
\begin{equation}\label{almost_min}
+\sum_{j_1,\dots,j_n\notin A}tr(B_{j_{n-1}}^{j_n}\cdots B_{j_1}^{j_2}B_{i}^{j_1}\rho B_{i}^{j_1*} B_{j_1}^{j_2*}\cdots B_{j_{n-1}}^{j_n*})x_k\Big(\frac{B_{j_{n-1}}^{j_n}\cdots B_{j_1}^{j_2}B_{i}^{j_1}\rho B_{i}^{j_1*} B_{j_1}^{j_2*}\cdots B_{j_{n-1}}^{j_n*}}{tr(B_{j_{n-1}}^{j_n}\cdots B_{j_1}^{j_2}B_{i}^{j_1}\rho B_{i}^{j_1*} B_{j_1}^{j_2*}\cdots B_{j_{n-1}}^{j_n*})}\Big)
\end{equation}
Finally suppose that $x_i(\rho)\geq 0$ for every $\rho$ and every $i$. Then the last term in (\ref{almost_min}) is also nonnegative. The remaining terms consists of $P(H^A(\rho)\leq n):=P(\{\textrm{time of first visit to A, given } \rho\}\leq n)$. So $x_i(\rho)\geq P(H^A(\rho)\leq n)$ for all $n$. Hence,
\begin{equation}
x_i(\rho)\geq \lim_{n\to \infty} P(H^A(\rho)\leq n)= P(H^A(\rho)< \infty)=h_i^A(\rho)
\end{equation}

\qed

\begin{remark}
In the Introduction we noted that the quantum trajectories can be studied in terms of a process with values on $D(\mathcal{H})\times \mathbb{Z}$ (see eq. (\ref{qtapp})). Another way of visualizing this process is to simply consider the state space to be $\mathbb{Z}$ but with varying probabilities (all depending on a fixed initial density), the proof of Theorem \ref{minimality_th} being a typical example of such approach.
\end{remark}

In a similar way, it is a simple matter to prove:
\begin{pro}\label{theorem_onk}
For every density $\rho\otimes |i\rangle\langle i|$, the vector of mean hitting time $(k_i^A(\rho))_{i\in\mathbb{Z}}$ is the minimal nonnegative solution to the system
\begin{equation}
\left\{
\begin{array}{ll}
k_i(\rho)=0 & \textrm{ if } i\in A  \\
k_i(\rho)=1+\sum_{j\notin A} tr(B_i^j\rho B_i^{j*})k_j\Big(\frac{B_i^j\rho B_i^{j*}}{tr(B_i^j\rho B_i^{j*})}\Big)
 & \textrm{ if } i\notin A
\end{array} \right.
\end{equation}
\end{pro}

\section{A class of channels}\label{zxsec5}

We are interested in nearest neighbor OQRWs,
\begin{equation}
\Phi(\rho)=\sum_{i\in\mathbb{Z}}\Big(B\rho_{i+1}B^*+C\rho_{i-1}C^*\Big)\otimes |i\rangle\langle i|,\;\;\;\rho=\sum_{i\in\mathbb{Z}}\rho_i\otimes|i\rangle\langle i|,
\end{equation}
where $B, C$ are normal, commuting matrices. In this section we will use the letters $B,C$ instead of the usual $L$, $R$ for nearest neighbor walks, as a reminder that we are assuming $BC=CB$. Due to commutativity, $B$ and $C$ are simultaneously diagonalizable via a unitary change of coordinates $U$ \cite{hj1} so that $B=UD_BU^*$, $C=UD_CU^*$, $D_B$, $D_C$ diagonal and we can write, for instance,
\begin{equation}
tr(BC\rho C^*B^*)=tr(UD_BD_CU^*\rho UD_B^*D_C^*U^*)=tr(D_BD_CU^*\rho UD_B^*D_C^*)=tr(D_B^*D_BD_C^*D_CU^*\rho U)
\end{equation}
and, more generally,
\begin{equation}\label{gen_comuteq_an}
tr(B^{l_1}C^{r_1}\cdots B^{l_n}C^{r_n}\rho C^{r_n*}B^{l_n*}\cdots C^{r_1*}B^{l_1*})=tr((D_B^*D_B)^{\sum_i l_i}(D_C^*D_C)^{\sum_i r_i}U^*\rho U)
\end{equation}
We conclude that the probability depends only on the number of times one moves left and right, and not on a particular sequence of $B$'s and $C$'s.

\medskip

{\bf Probability formula.} Given a state $\rho_a\otimes |a\rangle\langle a|$, we note that the probability of the walk to reach site $|x\rangle$ at time $n$ equals
\begin{equation}
p_x^{(n)}(\rho_a)=\sum_r M_n^r(a,x)tr((D_B^*D_B)^{n-r}(D_C^*D_C)^{r}U^*\rho_a U),\;\;\; n\geq 0,\;x\in\mathbb{Z},
\end{equation}
where $M_n^r(a,x)$ is the number of paths $(s_0,\dots, s_n)$, $s_0=a$, $s_n=x$ and having exactly $r$ rightward steps. Letting $l=n-r$ we get $r-l=x-a$ so we get $r=\frac{1}{2}(n+x-a)$ and $l=\frac{1}{2}(n-x+a)$ so
\begin{equation}\label{pformula1}
p_x^{(n)}(\rho_a)={n\choose \frac{1}{2}(n+x-a)}tr((D_B^*D_B)^{\frac{1}{2}(n-x+a)}(D_C^*D_C)^{\frac{1}{2}(n+x-a)}U^*\rho_a U),\;\;\; n\geq 0,\;x\in\mathbb{Z}
\end{equation}
Above we note that if $\frac{1}{2}(n+x-a)$ is not a positive integer then the probability equals zero.

\medskip

{\bf Hitting time formula.} As in previous sections, let $b_n(\rho_a;x)$ be the probability of first arrival at site $x$ occurring at time $n$, starting at $\rho_a\otimes|a\rangle\langle a|$. In case $a=0$, this is easily calculated by using the hitting time theorem \cite{grimmett}, which we state here for convenience of the reader.

\begin{theorem}[\cite{grimmett}, Section 3.10] The probability $f_b(n)$ that a random walk $S$ hits the point $b$ for the first time at the $n$-th step, having started from $0$, satisfies
\begin{equation}
f_b(n)=\frac{|b|}{n}P(S_n=b),\;\;\;n\geq 1.
\end{equation}
\end{theorem}
Then, together with probability formula (\ref{pformula1}) we have for $B,C$ commuting matrices,
\begin{equation}\label{unifiqf}
b_n(\rho_0;x)=\frac{|x|}{n}p_x^{(n)}=\frac{|x|}{n}{n\choose \frac{1}{2}(n+x)}tr((D_B^*D_B)^{\frac{1}{2}(n-x)}(D_C^*D_C)^{\frac{1}{2}(n+x)}U^*\rho_0 U),\;\;\; n\geq |x|,\;x\in\mathbb{Z}
\end{equation}

\begin{remark} Formula (\ref{unifiqf}) is a basic expression for the calculation of probabilities of nearest neighbor OQRWs on $\mathbb{Z}$ given by commuting matrices and, by such expression, we may assume $B$ and $C$ are diagonal by making a change of coordinates on $\rho_0$.
\end{remark}

\begin{remark}
Nearest neighbor classical processes are basic examples of the present setting, since in this case we may set $B_i^{j}$ as proper multiples of the identity (as in Example \ref{clbipr}, also see \cite{attal}). Conversely, consider a nearest neighbor OQRW with commuting matrices $B$ and $C$ (which we may assume are diagonal). Then the probability calculation above resembles the one for a classical process, but the precise resulting expression depends on the initial density matrix: for a given $\rho_0$, one needs the diagonal entries of $U^*\rho_0U$ in order to calculate $b_n(\rho_0;x)$.
\end{remark}

Now we illustrate how (\ref{unifiqf}) may be calculated in terms of the entries of $B$, $C$ and $\rho_0$. The following two examples have been previously discussed by N. Konno and H. Yoo \cite{konno}.

\begin{example}\label{konno1ex}
Let
$$B=\begin{bmatrix} 1 & 0 \\ 0 & \sqrt{p} \end{bmatrix},\;\;\;C=\begin{bmatrix} 0 & 0 \\ 0 & \sqrt{q} \end{bmatrix}$$
Then, for a initial matrix $\rho_0\otimes|0\rangle\langle 0|$, $\rho_0=diag(a,b)$,
\begin{equation}
p_x^{(n)}={n\choose \frac{1}{2}(n+x)}tr\Bigg(\begin{bmatrix}1 & 0 \\ 0 & p\end{bmatrix}^{\frac{1}{2}(n-x)}\begin{bmatrix}0 & 0 \\ 0 & q\end{bmatrix}^{\frac{1}{2}(n+x)}\begin{bmatrix}a & 0 \\ 0 & b\end{bmatrix}\Bigg)
\end{equation}
Then if $x=-n$ the above expression becomes
\begin{equation}
p_x^{(n)}=tr\Big(\begin{bmatrix}1 & 0 \\ 0 & p^n\end{bmatrix}\begin{bmatrix}a & 0 \\ 0 & b\end{bmatrix}\Big)=a+p^nb
\end{equation}
Otherwise, let $l=1/2(n-x)$ then $n-l=(n+x)/2$ and so
\begin{equation}
p_x^{(n)}=p_{n-2l}^{(n)}={n\choose n-l}tr\Bigg(\begin{bmatrix}1 & 0 \\ 0 & p\end{bmatrix}^{l}\begin{bmatrix}0 & 0 \\ 0 & q\end{bmatrix}^{n-l}\begin{bmatrix}a & 0 \\ 0 & b\end{bmatrix}\Bigg)={n\choose n-l}p^lq^{n-l}b
\end{equation}
i.e., this expression occurring whenever $x=n-2l$, $l=0,\dots,n$. This can be written as
\begin{equation}
p_x^{(n)}=a\delta_{x,-n}+b\sum_{l=0}^n{n\choose l}p^lq^{n-l}\delta_{x,n-2l},
\end{equation}
where $\delta_{x,y}=1$ if $x=y$ and equals zero otherwise.
\end{example}
\qee

\begin{example}\label{konno2ex}
Let
$$B=\begin{bmatrix} a(\epsilon) & \epsilon e^{i\theta} \\  \epsilon e^{i\theta} & a(\epsilon) \end{bmatrix},\;\;\;C=\begin{bmatrix} a(\epsilon) & -\epsilon e^{i\theta} \\  -\epsilon e^{i\theta} & a(\epsilon) \end{bmatrix},\;\;\;a(\epsilon)=\sqrt{1/2-\epsilon^2},\;\;\;\theta\in\mathbb{R}$$
where $\epsilon>0$ is such that $2\epsilon a(\epsilon)<1/2$. By \cite{konno}, these matrices are simultaneously diagonalizable so that
\begin{equation}
B^*B=U^*\begin{bmatrix} \lambda_+(\epsilon,\theta) & 0 \\ 0 & \lambda_-(\epsilon,\theta)\end{bmatrix}U,\;\;\;\;\;\;C^*C=U^*\begin{bmatrix} \lambda_-(\epsilon,\theta) & 0 \\ 0 & \lambda_+(\epsilon,\theta)\end{bmatrix}U,
\end{equation}
where $\lambda_{\pm}(\epsilon,\theta)=1/2\pm2\epsilon a(\epsilon)\cos(\theta)$ and
$$U=\frac{1}{\sqrt{2}}\begin{bmatrix} 1 & 1 \\ 1 & -1\end{bmatrix}$$
Then
$$p_x^{(n)}={n\choose \frac{1}{2}(n+x)}tr((D_B^*D_B)^{\frac{1}{2}(n-x)}(D_C^*D_C)^{\frac{1}{2}(n+x)}U^*\rho_0 U)$$
$$={n\choose \frac{1}{2}(n+x)}tr\Bigg(\begin{bmatrix} \lambda_+(\epsilon,\theta) & 0 \\ 0 & \lambda_-(\epsilon,\theta)\end{bmatrix}^{\frac{1}{2}(n-x)}\begin{bmatrix} \lambda_-(\epsilon,\theta) & 0 \\ 0 & \lambda_+(\epsilon,\theta)\end{bmatrix}^{\frac{1}{2}(n+x)}\begin{bmatrix} a & 0 \\ 0 & b\end{bmatrix}\Bigg)$$
\begin{equation}
={n\choose \frac{1}{2}(n+x)}tr\Bigg(\begin{bmatrix} a\lambda_+(\epsilon,\theta)^{\frac{1}{2}(n-x)}\lambda_-(\epsilon,\theta)^{\frac{1}{2}(n+x)} & 0 \\ 0 & b\lambda_-(\epsilon,\theta)^{\frac{1}{2}(n-x)}\lambda_+(\epsilon,\theta)^{\frac{1}{2}(n+x)}\end{bmatrix}\Bigg)
\end{equation}
Above note that $a=(U^*\rho_0 U)_{11}$ and $b=(U^*\rho_0 U)_{22}$. Again let $l=1/2(n-x)$ so we can write
\begin{equation}
p_x^{(n)}=p_{n-2l}^{(n)}={n\choose n-l}\Big(a\lambda_+(\epsilon,\theta)^{l}\lambda_-(\epsilon,\theta)^{n-l}+
                        b\lambda_-(\epsilon,\theta)^{l}\lambda_+(\epsilon,\theta)^{n-l}\Big)
\end{equation}
i.e., this expression occurring whenever $x=n-2l$, $l=0,\dots,n$.
\end{example}
\qee

\begin{example} Let
$$B=\frac{1}{4}\begin{bmatrix} 1+\sqrt{2} & 1-\sqrt{2} \\ 1-\sqrt{2} & 1+\sqrt{2}\end{bmatrix},\;\;\;C=\frac{1}{4}\begin{bmatrix} \sqrt{3}+\sqrt{2} & \sqrt{3}-\sqrt{2} \\ \sqrt{3}-\sqrt{2} & \sqrt{3}+\sqrt{2}\end{bmatrix}$$
Then $B^*B+C^*C=I$, $BC=CB$. Suppose $\Phi(\rho)=\sum_{i\in\mathbb{Z}}( B\rho_{i+1}B+C\rho_{i-1}C)\otimes|i\rangle\langle i|$, and take the initial state $\rho^{(0)}=\rho_0\otimes |0\rangle\langle 0|$, $\rho_0=(\rho_{ij})$. Then the probability of presence in site $|-1\rangle$ is $tr(B\rho_0B)=\frac{1}{8}(3\rho_{11}-\rho_{12}-\rho_{21}+3\rho_{22})$ and the probability of presence in site $|1\rangle$ is $tr(C\rho_0C)=\frac{1}{8}(5\rho_{11}+\rho_{12}+\rho_{21}+5\rho_{22})$. Also,
\begin{equation}
B=U\begin{bmatrix} \frac{1}{\sqrt{4}} & 0 \\ 0 & \frac{1}{\sqrt{2}}\end{bmatrix}U^*,\;\;\;C=U\begin{bmatrix} \frac{\sqrt{3}}{\sqrt{4}} & 0 \\ 0 & \frac{1}{\sqrt{2}}\end{bmatrix}U^*,\;\;\;U=\frac{1}{\sqrt{2}}\begin{bmatrix} 1 & 1 \\ 1 & -1\end{bmatrix}
\end{equation}
If $\rho_0=diag(1/3,2/3)$ then
\begin{equation}
p_{|2\rangle}^{(6)}={6\choose \frac{1}{2}(6+2)}tr(D_B^{6-2}D_C^{6+2}U^*\rho_0 U)=15tr(D_B^{4}D_C^{8}U\rho_0 U)
\end{equation}
But
\begin{equation}
D_B^4D_C^8U\rho_0U=\begin{bmatrix} \frac{81}{8192} & -\frac{27}{8192} \\ -\frac{1}{384} & \frac{1}{128} \end{bmatrix},
\end{equation}
so
\begin{equation}
p_{|2\rangle}^{(6)}=15.\frac{145}{8192}\approx0.2655029
\end{equation}
\end{example}
\qee

\section{Hitting time calculation: examples}\label{zxsec6}

All of the following examples present a common thread: these are motivated by some classical result, the transition probabilities have a matrix dependence (density matrices being of greater interest to us) and make use of spectral information associated to the matrices for an OQRW. As for the open walks, the assumption of these being nearest-neighbor and with commuting normal matrices simplifies the actual calculations while still providing nontrivial outcomes. For the corresponding classical examples, see \cite{norris}.

\begin{remark}\label{remark_breakparts}
We will be interested in the eigenmatrices $\{\eta_i\}$ of maps of the form $\Phi_V(X)=VXV^*$ for some $V$. Then we perform hitting time calculations for these particular matrices and take linear combinations of the results. However we emphasize that, in general, not every eigenmatrix for $\Phi_V$ is a density matrix (or even positive semidefinite). Because of this and due to the linearity of $h_i^A(\rho)$ with respect to $\rho$, we will be able to write
\begin{equation}
h_i^A(\rho)=h_i^A(\sum_j c_j\eta_j)=\sum_j c_j h_i^A(\eta_j),
\end{equation}
but in principle the individual terms $h_i^A(\eta_j)$ will have a quantum meaning only in the case $\eta_j$ is a (multiple of a) density matrix. In addition, it is instructive to observe what happens in the commutative case: from the probability formula (\ref{pformula1}), one considers a calculation of the form $tr(DU^*\rho U)$, with $D=diag(d_{11},d_{22})$. But this equals $d_{11}(U^*\rho U)_{11}+d_{22}(U^*\rho U)_{22}$. In other words, the probability is determined by $D$ and the diagonal entries of $\rho'=U^*\rho U$, so for calculation purposes, one only needs to consider the cases of $\rho$ such that $\rho'$ is diagonal: $\rho'=E_{11}$ or $E_{22}$ (projection on first and second coordinates, respectively) or convex combinations of those elements. The more general cases follows from such result.
\end{remark}

\begin{example} {\bf Walk on 4 sites: an open quantum version}. Here we revisit Example \ref{walk4sitecla}. Consider an OQRW on sites $i=1,2,3,4$. Denoting by $B_i^j$ the transition matrix from site $i$ to site $j$, we assume that for all $i$, $\sum_j B_i^{j*}B_i^j=I$. Suppose we allow nearest neighbor transitions only, given by
\begin{equation}
B_1^1=I,\;\;B_1^2=0,\;\;B_2^1=B_3^2=L, \;\;B_2^3=B_3^4=R,\;\;B_4^4=I,\;\;B_4^3=0,
\end{equation}
where we assume $L$ and $R$ are hermitian commuting matrices via a unitary change of coordinates $U$ so that $L=UD_LU^*$ and $R=UD_RU^*$, $D_L$ and $D_R$ diagonal. The more general case of normal commuting matrices is analogous. Once again we would like to calculate probability of ever reaching site $4$. Recalling the definition of $\pi(i;A)$ (paragraph preceding equation (\ref{umadef_imp})), we condition on the first step to get
$$
h_2^4(\rho_2^{(0)})=tr(B_2^1\rho_2^{(0)}B_2^{1*})\sum_{C\in\pi(1;4)} \frac{tr(CB_2^1\rho_2^{(0)}B_2^{1*}C^*)}{tr(B_2^1\rho_2^{(0)}B_2^{1*})}+tr(B_2^3\rho_2^{(0)}B_2^{3*})\sum_{D\in\pi(3;4)} \frac{tr(DB_2^3\rho_2^{(0)}B_2^{3*}D^*)}{tr(B_2^3\rho_2^{(0)}B_2^{3*})}$$
\begin{equation}\label{eqnor1a}
=tr(B_2^3\rho_2^{(0)}B_2^{3*})\sum_{D\in\pi(3;4)} \frac{tr(DB_2^3\rho_2^{(0)}B_2^{3*}D^*)}{tr(B_2^3\rho_2^{(0)}B_2^{3*})}=tr(B_2^3\rho_2^{(0)}B_2^{3*})h_3^4(\rho_3^{(1)}),
\end{equation}
where
\begin{equation}
\rho_3^{(1)}=\frac{B_2^3\rho_2^{(0)}B_2^{3*}}{tr(B_2^3\rho_2^{(0)}B_2^{3*})}
\end{equation}
Here we used the fact that a move from 1 to 4 is impossible. Now, since 4 is an absorbing state (i.e. $B_4^i=0$, $i\neq 4$), we have
$$
h_3^4(\rho_3^{(1)})=tr(B_3^4\rho_3^{(1)}B_3^{4*})\sum_{F\in\pi(4;4)} \frac{tr(FB_3^4\rho_3^{(1)}B_3^{4*}F^*)}{tr(B_3^4\rho_3^{(1)}B_3^{4*})}+tr(B_3^2\rho_3^{(1)}B_3^{2*})\sum_{G\in \pi(2;4)} \frac{tr(GB_3^2\rho_3^{(1)}B_3^{2*}G^*)}{tr(B_3^2\rho_3^{(1)}B_3^{2*})}
$$
$$=tr(B_3^4\rho_3^{(1)}B_3^{4*})+tr(B_3^2\rho_3^{(1)}B_3^{2*})\sum_{G\in \pi(2;4)} \frac{tr(GB_3^2\rho_3^{(1)}B_3^{2*}G^*)}{tr(B_3^2\rho_3^{(1)}B_3^{2*})}$$
\begin{equation}\label{spec_idea}
=tr\Big(B_3^4\frac{B_2^3\rho_2^{(0)}B_2^{3*}}{tr(B_2^3\rho_2^{(0)}B_2^{3*})}B_3^{4*}\Big)+tr\Big(B_3^2\frac{B_2^3\rho_2^{(0)}B_2^{3*}}{tr(B_2^3\rho_2^{(0)}B_2^{3*})}B_3^{2*}\Big)\sum_{G\in \pi(2;4)} \frac{tr(GB_3^2B_2^3\rho_2^{(0)}B_2^{3*}B_3^{2*}G^*)}{tr(B_3^2B_2^3\rho_2^{(0)}B_2^{3*}B_3^{2*})}
\end{equation}
Let $A=B_3^2B_2^3=LR$ and let $\Phi_A(X)=AXA^*$. Let $\{\eta_i\}$ a basis for $M_2(\mathbb{C})$ consisting of eigenstates for $\Phi_A$ and write $\Phi_A(\eta_i)=\lambda_i\eta_i$. So now we specialize to the case in which the initial state is an eigenstate (see Remark \ref{remark_breakparts}). Suppose $\rho_2^{(0)}=\eta_1$, then $B_3^2B_2^3\rho_2^{(0)}B_2^{3*}B_3^{2*}=\lambda_1 \rho_2^{(0)}$. As a consequence, (\ref{spec_idea}) becomes
$$h_3^4(\rho_3^{(1)})=tr\Big(B_3^4\frac{B_2^3\rho_2^{(0)}B_2^{3*}}{tr(B_2^3\rho_2^{(0)}B_2^{3*})}B_3^{4*}\Big)+tr\Big(B_3^2\frac{B_2^3\rho_2^{(0)}B_2^{3*}}{tr(B_2^3\rho_2^{(0)}B_2^{3*})}B_3^{2*}\Big)\sum_{G\in \pi(2;4)}tr(G\rho_2^{(0)}G^*)$$
\begin{equation}
=tr\Big(B_3^4\frac{B_2^3\rho_2^{(0)}B_2^{3*}}{tr(B_2^3\rho_2^{(0)}B_2^{3*})}B_3^{4*}\Big)+\frac{\lambda_1}{tr(B_2^3\rho_2^{(0)}B_2^{3*})}h_2^4(\rho_2^{(0)}).
\end{equation}
Now replace the above equation
into
\begin{equation}
h_2^4(\rho_2^{(0)})=tr(B_2^3\rho_2^{(0)}B_2^{3*})h_3^4(\rho_3^{(1)}),
\end{equation}
so we get
\begin{equation}
h_2^4(\rho_2^{(0)})=tr(B_3^4B_2^3\rho_2^{(0)}B_2^{3*}B_3^{4*})+\lambda_1h_2^4(\rho_2^{(0)}),
\end{equation}
from which we conclude
\begin{equation}\label{qresp1}
h_2^4(\rho_2^{(0)})=\frac{tr(B_3^4B_2^3\rho_2^{(0)}B_2^{3*}B_3^{4*})}{1-\lambda_1}=\frac{tr(R^2\rho_2^{(0)}R^{2*})}{1-\lambda_1}
\end{equation}
Note that the calculation above implies, as a particular case, the result from the classical matrix (\ref{clmatrix}) in Example \ref{walk4sitecla}: in this case there is no density matrix dependence and by letting $B_2^{3}=B_3^{4}=1/\sqrt{2}$, we get $\lambda_1=1/4$ and (\ref{qresp1}) equals $1/3$, as expected.
\end{example}
\qee

The following result concerns open quantum versions of two well-known classical chains. The proofs can be seen in the Appendix and are variations of the classical proofs.

\begin{theorem}\label{gambbd_ex}
a) {\bf Gambler's ruin.} Consider a nearest neighbor OQRW on $\mathbb{Z}$ with transition matrices $L$ and $R$ which commute. Let $\Phi_{L}(\rho)=L\rho L^*$ and $\Phi_{R}(\rho)=R\rho R^*$ with eigenmatrices $\Phi_{L}(\eta_j)=\lambda_{j}\eta_j$ and $\Phi_{R}(\eta_j)=\mu_{j}\eta_j$. Let $A=\{0\}$ and for every $\rho=\eta_j$ consider the system of equations
\begin{equation}
\left\{
\begin{array}{ll}
h_0^A(\rho)=1 &   \\
h_i^A(\rho)=p(\rho)h_{i+1}^A\Big(\frac{R\rho R^*}{tr(R\rho R^*)}\Big)+q(\rho)h_{i-1}^A\Big(\frac{L\rho L^*}{tr(L\rho L^*)}\Big) & i=1,2,\dots
\end{array} \right.
\end{equation}
where above we let $p(\rho)=tr(R\rho R^*)$, $q(\rho)=tr(L\rho L^*)$. If $\lambda_j\geq\mu_j$ then $h_i^A(\eta_j)=1$ for all $i$. If $\lambda_j<\mu_j$ then the minimal solution is
\begin{equation}
h_i^A(\eta_j)=\Big(\frac{\lambda_j}{\mu_j}\Big)^i
\end{equation}

b) {\bf Birth-and-death chain.} Consider a nearest neighbor OQRW on $\mathbb{Z}$ such that each site $|i\rangle$ has transition matrices $L_i$ and $R_i$, and assume $L_iR_j=R_jL_i$ for all $i,j$. Let $\Phi_{L_i}(\rho)=L_i\rho L_i^*$ and $\Phi_{R_i}(\rho)=R_i\rho R_i^*$ with eigenmatrices $\Phi_{L_i}(\eta_{j})=\lambda_{i;j}\eta_j$ and $\Phi_{R_i}(\eta_j)=\mu_{i;j}\eta_j$. Let $A=\{0\}$ and for every $\rho=\eta_j$ consider the system of equations
\begin{equation}
\left\{
\begin{array}{ll}
h_0^A(\rho)=1 &   \\
h_i^A(\rho)=p_i(\rho)h_{i+1}^A\Big(\frac{R_i\rho R_i^*}{tr(R_i\rho R_i^*)}\Big)+q_i(\rho)h_{i-1}^A\Big(\frac{L_i\rho L_i^*}{tr(L_i\rho L_i^*)}\Big),
 & i=1,2,\dots
\end{array} \right.
\end{equation}
Above we let $p_i(\rho)=tr(R_i\rho R_i^*)$, $q_i(\rho)=tr(L_i\rho L_i^*)$. Let
\begin{equation}
\gamma_i(\eta_j):=\frac{\lambda_{i;j}\lambda_{i-1;j}\cdots \lambda_{1;j}}{\mu_{i;j}\mu_{i-1;j}\cdots \mu_{1;j}}
\end{equation}
If $\sum_{i=0}^\infty\gamma_i(\eta_j)=\infty$ then $h_i^A(\eta_j)=1$ for all $i$. If $\sum_{i=0}^\infty\gamma_i(\eta_j)<\infty$ then for all $j$,
\begin{equation}
h_i(\eta_j)=\frac{\sum_{k=i}^\infty\gamma_k(\eta_j)}{\sum_{k=0}^\infty \gamma_k(\eta_j)},\;\;\; i=1,2,\dots
\end{equation}
\end{theorem}

Concerning the gambler's ruin, condition $\mu=\lambda$ implies that even if you find a ``fair casino", you are certain to end up broke. The calculation seen in the proof is quite similar to the corresponding classical version, except that the transition probabilities are induced by matrices $L$ and $R$ for all sites. Also it should be clear that now one may consider arbitrary density matrices $\rho$. Therefore we have a dependence on the basis $\eta_j$ and, correspondingly, on the eigenvalues $\lambda_{j}, \mu_{j}$, and these may now be considered in linear superposition. For instance, let $\eta_x$ and $\eta_y$ be two eigenstates. Then
$$h_i^A(\alpha\eta_x+(1-\alpha)\eta_y)=\sum_{D\in\pi(i;A)}tr(D[\alpha\eta_x+(1-\alpha)\eta_y]D^*)=\alpha \sum_{D\in\pi(i;A)}tr(D\eta_x D^*)+(1-\alpha)\sum_{D\in\pi(i;A)}tr(D\eta_y D^*)$$
\begin{equation}\label{outcomm}
=\alpha h_i^A(\eta_x)+(1-\alpha)h_i^A(\eta_y)
\end{equation}
so we have a combination of two previously known answers. Each term in (\ref{outcomm}) can be seen as a contribution to the final outcome and on each of them we may have a distinct situation. For instance, if $\mu_x<\lambda_x$, $\mu_y>\lambda_y$ are such that $\Phi_L(\eta_x)=\lambda_x\eta_x$, $\Phi_R(\eta_x)=\mu_x\eta_x$ and $\Phi_L(\eta_y)=\lambda_y\eta_y$, $\Phi_R(\eta_y)=\mu_y\eta_y$ then $h_i^A(\eta_x)=1$ for all $i$, $h_i^A(\eta_y)=(\lambda_y/\mu_y)^i$ and so
\begin{equation}
h_i^A(\alpha\eta_x+(1-\alpha)\eta_y)=\alpha +(1-\alpha)\Big(\frac{\lambda_y}{\mu_y}\Big)^i
\end{equation}
Hence we have concluded that the gambler's ruin is certain only in the case that a density matrix, when written in the basis given by the common eigenmatrices $\{\eta_i\}$ of $L$ and $R$, does not contain any $\eta_i$ such that $\mu_i\geq\lambda_i$. In other words: if, for some contribution, moving away from ruin has a larger weight than moving left, then ruin is not certain (also see Remark \ref{remark_breakparts}).

\medskip

Concerning the birth-and-death chain we have that, as in the classical case, $h_i<1$ implies that the population survives with positive probability. Therefore we see that this survival depends on the $\gamma_i$ which, in turn, depends on the $p_i, q_i$ and $\eta_j$. In a similar way as in the gambler's ruin, we have that the dependence on $\eta_j$, not seen in the classical birth-and-death chain, can be interpreted as an extra degree of freedom of the walk just described. As a particular example let $\eta_x$ and $\eta_y$ be two eigenstates and write
\begin{equation}\label{bd_sol1}
h_i^A(\alpha\eta_x+(1-\alpha)\eta_y)=\alpha h_i^A(\eta_x)+(1-\alpha)h_i^A(\eta_y)
\end{equation}
Let $\eta_x$ be such that $\sum_{i=0}^\infty\gamma_i(\eta_x)=\infty$, so we have $h_i^A(\eta_x)=1$. And let $\eta_y$ be such that $\sum_{i=0}^\infty\gamma_i(\eta_y)<\infty$, so we have the solution
\begin{equation}
h_i(\eta_y)=\frac{\sum_{k=i}^\infty\gamma_k(\eta_y)}{\sum_{k=0}^\infty \gamma_k(\eta_y)},\;\;\;\forall i=1,2,\dots
\end{equation}
Then (\ref{bd_sol1}) becomes
\begin{equation}
h_i^A(\alpha\eta_x+(1-\alpha)\eta_y)=\alpha+(1-\alpha)\frac{\sum_{k=i}^\infty\gamma_k(\eta_y)}{\sum_{k=0}^\infty \gamma_k(\eta_y)}
\end{equation}
In an analogous way as in the gambler's ruin we have: if for some contribution survival is possible, then population extinction is not certain to occur.

\subsection{A theorem on potential theory}

Let $D=\{-a,-a+1,\dots,a-1,a\}$, $a\in\mathbb{Z}$ and $\rho_0$ be an initial density state on $0$. We are interested in the average cost of going from $0$ to the boundary of $D$, that is $\partial D=\{-a,a\}$ (for simplicity we discuss the problem on $\mathbb{Z}$, but one can extend the discussion to $\mathbb{Z}^d$ if desired). In the classical theory, we can write such cost in terms of a potential
\begin{equation}
\phi_i^c=E_i\Big(\sum_{n<T} c(X_n)+f(X_T)1_{T<\infty}\Big)
\end{equation}
where $c$ denotes a cost inside $D$, $f$ is the cost at $\partial D$ (i.e, a boundary value) and $T$ is the hitting time of $\partial D$. We can study an OQRW analog of this cost problem. To do this, we assume $(c(i):i\in D)$ and $(f(i):i\in\{-a,a\})$ are nonnegative. Then if $\{X_i\}_{i\geq 1}$ denotes a particular position on $\mathbb{Z}$ via a quantum trajectories process, define
\begin{equation}\label{potdef1}
\phi_i^c(\rho):=\sum_{X_0=i,X_1,X_2,\dots\in(-a,a), X_r\in\{-a,a\}}[c(X_0)+c(X_1)+c(X_2)+\cdots+c(X_{r-1})+f(X_r)]P_i(X_1,\dots,X_r;\rho)
\end{equation}
where
\begin{equation}
P_i(X_1,\dots,X_r;\rho):=tr(V_r\cdots V_1\rho V_1^*\cdots V_r),\;\;\;\;\;\;V_i=V_{X_{i-1}X_i}=B_{X_{i-1}}^{X_i}
\end{equation}
Expression (\ref{potdef1}) means that we sum the costs for all the possible ways one can go from site $i$, with initial density matrix $\rho$ and reach the border (which in the one-dimensional case described is just the set $\{a,-a\}$), multiplied by the probability for the corresponding paths to occur. Let
\begin{equation}
\phi_i(\rho|X_1=j):=\phi_j\Big(\frac{B_i^j\rho B_ i^{j*}}{tr(B_i^j\rho B_ i^{j*})}\Big)
\end{equation}
Then for $i\in D$,
\begin{equation}\label{pot_eq1}
\phi_i(\rho)=c(i)+\sum_{j\in I}p_{ij}(\rho)\phi_i(\rho|X_1=j)=c(i)+\sum_{j\in I}p_{ij}(\rho)\phi_j\Big(\frac{B_i^j\rho B_i^{j*}}{tr(B_i^j\rho B_i^{j*})}\Big),\;\;\;p_{ij}(\rho)=tr(B_i^j\rho B_i^{j*})
\end{equation}
and $\phi_i(\rho)=f(i)$ in $\partial D$. Above we chose $I=\mathbb{Z}$ (set $\mathbb{Z}^d$ for the general case). In an analogous way as in the hitting time problems discussed previously, if $\rho$ is 1-dimensional then we recover the classical expression: if $B_i^j=\sqrt{p_{ij}}I$ we get $tr(B_i^j\rho_i B_i^{j*})=p_{ij}$ and
\begin{equation}\label{norris_matrix_eq}
\phi_i=c(i)+\sum_{j\in I}p_{ij}\phi_i|_{X_1=j}=c(i)+\sum_{j\in I}p_{ij}\phi_j
\end{equation}
Equation (\ref{norris_matrix_eq}) is just the matrix equation $\phi=c+P\phi$. We define the {\bf open quantum version} of $\phi=c+P\phi$ to be
\begin{equation}
\phi_i(\rho)=c(i)+\sum_{j\in I}p_{ij}(\rho)\phi_j\Big(\frac{B_i^j\rho B_i^{j*}}{tr(B_i^j\rho B_i^{j*})}\Big),\;\;\;i\in D
\end{equation}
The proof of the following can be seen in the Appendix.
\begin{theorem}\label{teo_pot11}
Suppose that $\{c(i):i\in D\}$ and $\{f(i):i\in\partial D\}$ are nonnegative. Let
\begin{equation}
\phi_i(\rho):=\sum_{X_0=i,X_1,X_2,\dots\in(-a,a), X_r\in\{-a,a\}}[c(X_0)+c(X_1)+c(X_2)+\cdots+c(X_{r-1})+f(X_r)]P(X_1,\dots,X_r;\rho)
\end{equation}
Then for every density $\rho$ the following holds. a) The potential $\phi=\{\phi_i:i\in I\}$ satisfies
\begin{equation}\label{cond_itema}
\phi_i(\rho)=c(i)+\sum_{j\in I}p_{ij}(\rho)\phi_i(\rho|X_1=j)=c(i)+\sum_{j\in I}tr(B_i^j\rho B_i^{j*})\phi_j\Big(\frac{B_i^j\rho B_i^{j*}}{tr(B_i^j\rho B_i^{j*})}\Big),
\end{equation}
and $\phi=f$ in $\partial D$. b) If $\{\psi_i:i\in I\}$ satisfies the open quantum version of $\psi\geq c+P\psi$ in $D$ and $\psi\geq f$ in $\partial D$ and $\psi_i\geq 0$ for all $i$ then $\psi_i\geq \phi_i$ for all $i$. c) If $P_i(T<\infty)=1$ for all $i$ then (\ref{cond_itema})  has at most one bounded solution.
\end{theorem}

\begin{example}
Let $P=P(\rho)$ be some matrix operator which depends on a density $\rho$ and let $\phi=(\phi_i)$ be a bounded nonnegative function such that $P(\rho)\phi=\phi$ on $D$, a certain collection of sites. Assume the cost function inside $D$ equals zero, i.e., $c\equiv 0$. If the probability of ever reaching $\partial D$ is 1 then $\phi$ is entirely determined by its value on the boundary. In fact, let $f$ be the restriction of $\phi$ to $\partial D$. Then
\begin{equation}
\phi_i(\rho)=\left\{
\begin{array}{ll}
\sum_{j\in I}tr(B_i^j\rho B_i^{j*})\phi_j\Big(\frac{B_i^j\rho B_i^{j*}}{tr(B_i^j\rho B_i^{j*})}\Big) & \textrm{ if } i\in D  \\
f(i) & \textrm{ if } i\in\partial D
\end{array} \right.
\end{equation}
As $f$ is bounded the function $i\mapsto \phi_i^0(\rho)$, where
\begin{equation}\label{ex_sol1}
\phi_i^0(\rho):=\sum_{X_0=i,X_1,X_2,\dots\in(-a,a), X_r\in\{-a,a\}}f(X_r)P_i(X_1,\dots,X_r;\rho)
\end{equation}
is bounded and therefore, since reaching the boundary is assumed to be certain, it is the unique nonnegative bounded solution of (\ref{cond_itema}) for $c=0$.
As an example, consider a walk on $\mathbb{Z}^2$, write $e_1=(1,0)$, $e_2=(0,1)$ and let
\begin{equation}
P(\rho)v(i)=v(i+e_1)p_{i+e_1,i}(\rho)+v(i-e_1)p_{i-e_1,i}(\rho)+v(i+e_2)p_{i+e_2,i}(\rho)+v(i-e_2)p_{i-e_2,i}(\rho)
\end{equation}
This describes a nearest neighbor walk which is dependent on a particular density $\rho$. We can define \begin{equation}
\Delta_\Phi v_i(\rho)=\sum_{j\in I}tr(B_i^j\rho B_i^{j*})v_j\Big(\frac{B_i^j\rho B_i^{j*}}{tr(B_i^j\rho B_i^{j*})}\Big)-v_i(\rho)
\end{equation}
as a kind of generalized Laplacian. If the dependence on $\rho$ vanishes and the matrix represents a symmetric walk, we recover the classical case and then the problem
\begin{equation}\label{dirichlet_p}
\left\{
\begin{array}{ll}
\Delta_\Phi v_i(\rho)=0 & \textrm{ if } i\in D \\
v_i(\rho)=f(i) & \textrm{ if } i\in\partial D
\end{array} \right.
\end{equation}
reduces to the usual Dirichlet problem. In the more general case, under the assumption that $\partial D$ is nonempty and that a walk reaches the boundary with probability 1, we have seen that (\ref{cond_itema}) has at most one bounded solution. If $f$ is bounded, the solution is (\ref{ex_sol1}). Hence if $\partial D$ is nonempty and (\ref{ex_sol1}) is bounded, then it is the unique nonnegative bounded solution of (\ref{dirichlet_p}). For more on the classical version of this result, see \cite{bremaud}.

\end{example}
\qee

\section{Conclusions and open questions: hitting times for nonhomogeneous quantum Markov chains}\label{sec_htnhqmc}

In this work we have studied a class of quantum Markov chains induced by transition matrices. We have also described such objects in terms of open quantum random walks and, together with the results on ergodicity, we hope this will encourage further research on dynamical aspects of these operators. Among basic open questions is the problem of establishing results for QTMs which are not unital, that is, such that the associated OQRW does not preserve the identity. In this case we note that the property that all entries of a given stationary state are equal (a fact which allows certain calculations performed in our adaptation for the unital case) cannot be used. Many of the questions made here make sense for QTMs acting on an infinite dimensional space, but in principle describing and solving these problems would require a treatment which is different than the one used here.

\medskip

Concerning the theory described in \cite{saloff} one might be interested in studying quantum adaptations of other properties presented there, particularly aspects appearing in processes that involve groups. Also according to this work, if every stochastic matrix $Q_i$ is reversible with respect to a given distribution, i.e.,
\begin{equation}
\pi(x)Q_i(x,y)=\pi(y)Q_i(y,x),\;\;\;\forall i,
\end{equation}
then the condition $\sigma_2(Q_i)<1$ for each $i$ is equivalent to the fact that each $Q_i$ is irreducible and aperiodic. Then one may ask for an appropriate notion of reversibility for the setting presented in this work. We refer the reader to \cite{temme} for quantum versions of the detailed balance equation.

\medskip

In connection to other quantum contexts where hitting times have been previously discussed, it is an open question to understand how the structure presented here behaves when the channel given by OQRWs is slightly changed: recall that a density of the form $\rho=\sum_i \rho_i\otimes |i\rangle\langle i|$ has its form preserved under the action of an OQRW $\Phi$ (i.e., $\Phi(\rho)$ continues to be a linear combination of tensor product of positive matrices and projections operators). One may be interested in different kinds of densities and to inspect what happens to hitting times (and the particular case of recurrence) in these more general contexts.

\medskip

As a final discussion, consider a classical, nonhomogeneous Markov chain $\{P_n\}$ acting on a finite state space $\Omega$. Suppose $A\subset\Omega$ is a transient class, $|A|=d$ and let $P_n^A$ denote the restriction of $P_n$ to $A\times A$. Write the transition matrix of this chain as
\begin{equation}
P_n=\begin{bmatrix}
P_n^A & C_n \\ 0 & I
\end{bmatrix}
\end{equation}
Then the distribution of the hitting time in the subset $\Omega\setminus A$ at time $n$, assuming an initial distribution $\mu_A$, can be written as
\begin{equation}
P(T>n)=P(X_i\in A,\;i\leq n)=\mu_A\Pi_{k=0}^{n-1}P_k^A1_d,\;\;\;n\geq 1,
\end{equation}
where $1_d=(1\; 1\cdots 1)^T$ is $d$-dimensional. The mean hitting time is related to the behavior of the product $P_{n_1}^AP_{n_2}^A\cdots$. Following \cite{platis}, we say that $A$ is a $\theta$-transient class at time $n\geq 0$ if all $i\in A$ at time $n$ leads to a state in $\Omega\setminus A$ with a transition probability at least equal to $\theta$. Then the authors have shown that if for all $n\geq 0$ $A$ is a $\theta$-transient class, then there exists an increasing sequence of positive integers $k_j$, $j=1,2,\dots$ and $0\leq\theta<1$ such that for all $j\in\mathbb{N}$, $\Vert P_{k_j}^A\cdots P_{k_{j+1}-1}^A\Vert\leq\theta<1$. The mentioned sequence provides a criterion for the finiteness of the mean hitting time for $\Omega\setminus A$ for a initial state in $A$, namely,
\begin{equation}
E(T)=\mu_A\Big[1_d+\sum_{n \geq 0}\Pi_{k=0}^n P_k^A1_d\Big]
\end{equation}
is finite, provided that
\begin{equation}
\lim_{m\to\infty}\sup (k_{m+1}-k_m)^{1/m}<\frac{1}{\theta}
\end{equation}
Then, in our context, a natural question is: can we obtain a similar criterion in order to describe mean hitting times for QTMs? The intermediate result which provides the criterion is the following, restated here for convenience.
\begin{pro}\cite{platis}
If $A$ is a transient class and $B$ is absorbing for all $n\geq 0$, then the series of terms $\Vert \Pi_{k=0}^n P_k^A\Vert$ converges if the power series $\sum_{n\geq 0} (k_{n+1}-k_n)\theta^n$ is convergent.
\end{pro}
One might expect a simple adaptation of the ideas seen in the mentioned work, but up to our knowledge such problem has not been investigated so far.

\section{Appendix: Proofs}

{\bf Proof of Proposition \ref{saloff_explicado}.} We have for $B=(B_{ij})$ QTM, for all $j$,
$$
\sum_{i=1}^n \Big\Vert B_{j}^{i*}B_{j}^i-\frac{I}{n}\Big\Vert_2^2=\sum_{i=1}^n \langle B_{j}^{i*}B_{j}^i,B_{j}^{i*}B_{j}^i\rangle_2-2\sum_{i=1}^n\langle B_{j}^{i*}B_{j}^i,\frac{I}{n}\rangle_2+n\langle\frac{I}{n},\frac{I}{n}\rangle_2$$
$$=\sum_{i=1}^ntr((B_{j}^{i*}B_{j}^i)^2)-2\langle I,\frac{I}{n}\rangle_2+\frac{1}{n}\langle I,I\rangle_2=\sum_{i=1}^ntr((B_{j}^{i*}B_{j}^i)^2)-\frac{1}{n}\langle I,I\rangle_2$$
\begin{equation}\label{im0}
=\sum_{i=1}^n tr((B_{j}^{i*}B_{j}^i)^2) -\frac{k}{n}
\end{equation}
With (\ref{im0}) in mind, we perform another calculation. Consider an orthonormal basis of eigenstates for $\Phi^*\Phi$ given by $\mathcal{F}=\{\eta_i\}_{i=1}^{N_\Phi}$, such that
\begin{equation}
\eta_1=\frac{1}{\sqrt{kn}}\Big(I\otimes |1\rangle\langle 1|+\cdots +I\otimes |n\rangle\langle n|\Big), \;\;\;I=I_k\in M_k(\mathbb{C}),
\end{equation}
where $I=I_k$ is the order $k$ identity matrix. Note that $\Vert \eta_1\Vert=1$. Recall that by Remark \ref{bigrem2}, whatever is the initial state $\rho$ on $\mathcal{H}\otimes\mathcal{K}$, the density $\Phi(\rho)$ is of the form $\sum_i\rho_i\otimes |i\rangle\langle i|$. This imposes a restriction on the kind of eigenstates present in $\mathcal{F}$. Define
\begin{equation}\label{rho1bom}
\rho_1:=\begin{bmatrix} I & 0 & \dots & 0\end{bmatrix}^T=\sum_{i=1}^{N_\Phi} d_i\eta_i,\;\;\;d_i\in\mathbb{C},
\end{equation}
with $N_\Phi$ as in Remark \ref{remark_4n2}. We have
\begin{equation}\label{im1}
\langle \Phi^*\Phi\rho_1,\rho_1\rangle_2=\langle\Phi^*\Phi\sum_{i} d_i\eta_i,\sum_{j} d_j\eta_j \rangle=\sum_{i,j} d_i\ov{d_j}\langle\Phi^*\Phi\eta_i,\eta_j\rangle=\sum_{i=1}^{N_\Phi} |d_i|^2\sigma_i^2
\end{equation}
Also,
\begin{equation}
\Phi(\rho_1)=\begin{bmatrix} B_{1}^1IB_{1}^{1*} & B_{1}^2IB_{1}^{2*} & \cdots & B_{1}^nIB_{1}^{n*}\end{bmatrix}^T,
\end{equation}
so
\begin{equation}\label{im2}
\langle \Phi(\rho_1),\Phi(\rho_1)\rangle_2=\Big\langle \begin{bmatrix} B_{1}^1IB_{1}^{1*} \\ B_{1}^2IB_{1}^{2*} \\ \vdots \\ B_{1}^nIB_{1}^{n*}\end{bmatrix},\begin{bmatrix} B_{1}^1IB_{1}^{1*} \\ B_{1}^2IB_{1}^{2*} \\ \vdots \\ B_{1}^nIB_{1}^{n*}\end{bmatrix}\Big\rangle_2=tr((B_{1}^{1*}B_{1}^1)^2)+tr((B_{1}^{2*}B_{1}^2)^2)+\cdots+tr((B_{1}^{n*}B_{1}^n)^2).
\end{equation}
Therefore on one hand we have
\begin{equation}
\sum_{i=1}^n tr((B_{1}^{i*}B_{1}^i)^2)=\sum_{i=1}^{N_\Phi} |d_i|^2\sigma_i^2,
\end{equation}
and on the other we obtained, by (\ref{im0}),
\begin{equation}
\sum_{i=1}^n \Big\Vert B_{1}^{i*}B_{1}^i-\frac{I}{n}\Big\Vert_2^2=\sum_{i=1}^n tr((B_{1}^{i*}B_{1}^i)^2) -\frac{k}{n}
\end{equation}
Finally, note that $d_1=\langle \rho_1,\eta_1\rangle=\frac{1}{\sqrt{kn}}tr(I)=\frac{k}{\sqrt{kn}}$ and so $|d_1|^2=\frac{k}{n}$. We can repeat an analogous reasoning where we define $\rho_2$, $\rho_3,\dots$ in a similar way as $\rho_1$ in (\ref{rho1bom}).

\qed

{\bf Proof of Theorem \ref{bigbigt}.} Assume that $\sigma=\max_{1,\dots, q}\sigma_2(Q_j)<1$. Let $\{B_i\}_{i=1}^\infty$ be a sequence of OQRWs such that
\begin{equation}
N_l=\#\{i\in\{1,\dots,l\}:B_i\in\mathcal{Q}\}
\end{equation}
tends to infinity with $l$. By (\ref{saloff_2}) we have that for every $j$,
\begin{equation}\label{part_facc1}
\Big(\sum_{i=1}^n \Big\Vert \mathcal{B}_{0,l}(i,j)\mathcal{B}_{0,l}(i,j)^*-\frac{I}{n}\Big\Vert_2^2\Big)^{1/2}\leq \sigma^{N_l}C(j,n)
\end{equation}
which tends to zero as $l\to\infty$. Conversely, assume that the pair $(\mathcal{Q},\rho_\pi)$ is ergodic. Then equation (\ref{linesequal}) holds for any sequence $(B_i)_1^\infty$ of QTMs with invariant measure $\rho_\pi$ such that $B_i\in\mathcal{Q}$ for infinitely many $i$'s. That is, the columns of the iterated product are becoming equal to $I/n$. By contradiction, assume that one of the $Q_i$, say $Q_1$ satisfies $\sigma_2(Q_1)=1$ and consider the following sequence of QTMs: $B_{2i+1}=Q_1$, $B_{2i}=Q_1^*$, $i=1,2,\dots$. Now we consider $Q_1Q_1^*$, for which $\sigma_2(Q_1)=1$ is an eigenvalue with algebraic and geometric multiplicity at least 2, i.e., $\mu_\Phi(1)=\gamma_\Phi(1)\geq 2$ (see Prop. \ref{wolf1}).

\smallskip

Now let $\Psi_r=(Q_1Q_1^*)^r$. It is clear that each $\Psi_r$ is a quantum channel with real spectrum in $[0,1]$, by the remarks preceding the statement of this theorem. By standard arguments such as the one seen in Novotn\'y et al. \cite{novotny,novotny2} (via Jordan blocks), the asymptotic behavior of $\Psi_r$ is determined by the peripheral spectrum, as contributions of eigenspaces associated to eigenvalues with norm less than 1 tend to disappear. Since $1$ is the only eigenvalue in the unit circle associated to $\Psi_r$, for all $r$, it is clear that the limit of $\Psi_r$ as $r\to\infty$ exists. The QTM $B=\lim_{r\to\infty}\Psi_r=\lim_{r\to\infty}(Q_1Q_1^*)^r$ is such that there must be two matrices $B(i,j)$ and $B(l,m)$ which are distinct. Indeed, let $\rho_0$ be an eigenstate of $B$ associated to eigenvalue 1. Suppose that $\rho_0$ is not the maximally mixed column. This assumption is possible since we have that the geometric multiplicity of 1 for $Q_1Q_1^*$ is at least 2. In particular, by writing $\rho_0=\sum_i \eta_i\otimes|i\rangle\langle i|$ we may assume there are $k,l$ such that $\eta_k\neq \eta_l$. Now the fact that $B(\rho_0)=\rho_0$, corresponds to the system of equations
\begin{equation}
B(1,1)\eta_1 B(1,1)^{*}+\cdots+B(1,n)\eta_n B(1,n)^{*}=\eta_1
\end{equation}
\begin{equation}
B(2,1)\eta_1 B(2,1)^{*}+\cdots+B(2,n)\eta_n B(2,n)^{*}=\eta_2
\end{equation}
$$\vdots$$
\begin{equation}
B(n,1)\eta_1 B(n,1)^{*}+\cdots+B(n,n)\eta_n B(n,n)^{*}=\eta_n
\end{equation}
If, on the contrary, all $B(i,j)$ are equal, then by considering the $k$-th and $l$-th equation we conclude that $\eta_k=\eta_l$, which is absurd. Therefore the QTM $B=\lim_{r\to\infty}\Psi_r$ is such that there must be two matrices $B(i,j)$ and $B(l,m)$ which are distinct.

\smallskip

Moreover, there must be a row in $B$ with two different entries, that is, two matrices $B(i,r)\neq B(i,s)$ for some $i,r,s$. In fact, suppose $i_1$ and $i_2$ are rows where we found two distinct elements of $B$, say, $B(i_1,j_1)$ and $B(i_2,j_2)$. If all entries of row $i_1$ are equal then these must be equal to the maximally mixed column. The same conclusion holds for row $i_2$. But then we would have $B(i_1,j_1)=B(i_2,j_2)$, which is absurd. We conclude there must be a row in $B$ with two different entries. Hence we are able to obtain $x,y,z$ such that
\begin{equation}
\lim_{r\to\infty} \mathcal{B}_{0,2r}(x,y)- \mathcal{B}_{0,2r}(x,z)\neq 0,
\end{equation}
as required.

\qed

{\bf Proof of Theorem \ref{gambbd_ex}}. a) Gambler's ruin. Let $A=\{0\}$ so that $h_i^A(\rho_i)$ is the ruin probability starting from $i$. Using Theorem \ref{minimality_th} we have the system of equations
\begin{equation}
\left\{
\begin{array}{ll}
h_0^A(\rho_i)=1 &   \\
h_i^A(\rho_i)=p(\rho_i)h_{i+1}^A\Big(\frac{B_i^{i+1}\rho_i B_i^{i+1*}}{tr(B_i^{i+1}\rho_i B_i^{i+1*})}\Big)+q(\rho_i)h_{i-1}^A\Big(\frac{B_i^{i-1}\rho_i B_i^{i-1*}}{tr(B_i^{i-1}\rho_i B_i^{i-1*})}\Big) & i=1,2,\dots
\end{array} \right.
\end{equation}
For instance if $i=2$, using expression (\ref{umadef_imp}) and recalling that $B_i^{i+1}=R$, $B_i^{i-1}=L$,
\begin{equation}
h_2^A(\rho)=tr(B_2^3\rho B_2^{3*})\sum_{D\in\pi(3;A)} \frac{tr(DB_2^3\rho B_2^{3*}D^*)}{tr(B_2^3\rho B_2^{3*})}+tr(B_2^1\rho B_2^{1*})\sum_{C\in\pi(1;A)} \frac{tr(CB_2^1\rho B_2^{1*}C^*)}{tr(B_2^1\rho B_2^{1*})}
\end{equation}
We will also write $q(\rho)=tr(L\rho L^*)$ and $p(\rho)=tr(R\rho R^*)$. Let $\Phi_L(X)=LXL^*$ and $\Phi_R(X)=RXR^*$. Then if $LR=RL$ we have $\Phi_R\Phi_L=\Phi_L\Phi_R$ so let $\{\eta_i\}$ be a basis for $M_2(\mathbb{C})$ consisting of eigenstates for $\Phi_L$ and $\Phi_R$ and write $\Phi_L(\eta_i)=\lambda_i\eta_i$, $\Phi_R(\eta_i)=\mu_i\eta_i$. Then we have
$$h_2^A(\eta_i)=tr(R\eta_i R^*)\sum_{D\in\pi(3;A)} \frac{tr(DB_2^3\eta_i B_2^{3*}D^*)}{tr(B_2^3\eta_i B_2^{3*})}+tr(L\eta_i L^*)\sum_{C\in\pi(1;A)} \frac{tr(CB_2^1\eta_i B_2^{1*}C^*)}{tr(B_2^1\eta_i B_2^{1*})}$$
$$tr(R\eta_i R^*)\sum_{D\in\pi(3;A)} \frac{tr(D\mu_i\eta_i D^*)}{\mu_i tr(\eta_i)}+tr(L\eta_i L^*)\sum_{C\in\pi(1;A)} \frac{tr(C\lambda_i\eta_i C^*)}{\lambda_i tr(\eta_i)}  $$
\begin{equation}
=p(\eta_i)h_{3}^A(\eta_i)+q(\eta_i)h_{1}^A(\eta_i)
\end{equation}
In a simpler notation, let $\eta$ be such that $\Phi_L(\eta)=\lambda\eta$ and $\Phi_R(\eta)=\mu\eta$, $\lambda,\eta\in\mathbb{C}$. Then for such choice the above equation becomes
\begin{equation}
h_i^A(\eta)=\mu h_{i+1}^A(\eta)+\lambda h_{i-1}^A(\eta).
\end{equation}
If $\mu \neq \lambda$ the recurrence relation has, for fixed $\eta$, a general solution
\begin{equation}
h_i^A(\eta)=B+C\Big(\frac{\lambda}{\mu}\Big)^i,\;\;\;\lambda=\lambda(\eta),\; \mu=\mu(\eta)
\end{equation}
Now, if $\mu<\lambda$ then the restriction $0\leq h_i\leq 1$ forces $C=0$ so $h_i=1$ for all $i$. That is, the ruin of the quantum gambler is certain. On the other hand, if $\mu>\lambda$ then we get the family of solutions
\begin{equation}
h_i^A(\eta)=\Big(\frac{\lambda}{\mu}\Big)^i+A\Big(1-\Big(\frac{\lambda}{\mu}\Big)^i\Big)
\end{equation}
For a nonnegative solution we must have $B\geq 0$ and so the minimal solution is
\begin{equation}
h_i^A(\eta)=\Big(\frac{\lambda}{\mu}\Big)^i
\end{equation}
Finally if $\mu=\lambda$, the recurrence relation has a general solution $h_i^A=B+Ci$ and again the restriction $0\leq h_i^A(\eta)\leq 1$ forces $C=0$, so $h_i^A(\eta)=1$ for all $i$.

\medskip

b) Birth-and-death chain. We write $p_i(\rho)=tr(R_i\rho R_i^*)$, $q_i(\rho)=tr(L_i\rho L_i^*)$. We still assume commutativity of the translation operators (moves to the left or right). Let $\Phi_{L_i}(\eta_j)=\lambda_{i;j}\eta_j$, $\Phi_{R_i}(\eta_j)=\mu_{i;j}\eta_j$. Then for any eigenstate $\eta_j$ we have
\begin{equation}\label{aux11}
h_i^A(\eta_j)=p_i(\eta_j)h_{i+1}^A\Big(\frac{R_i\eta_j R_i^*}{tr(R_i\eta_j R_i^*)}\Big)+q_i(\eta_j)h_{i-1}^A\Big(\frac{L_i\eta_j L_i^*}{tr(L_i\eta_j L_i^*)}\Big),
\end{equation}
thus implying, due to the eigenvalue conditions $p_i(\eta_j)=\mu_{ij}$, $q_i(\eta_j)=\lambda_{ij}$ and using the same calculation used in the gambler's ruin,
\begin{equation}
h_i^A(\eta_j)=\mu_{i;j}h_{i+1}^A(\eta_j)+\lambda_{i;j} h_{i-1}^A(\eta_j).
\end{equation}
Let $u_i=h_{i-1}^A-h_i^A$, then
\begin{equation}
\mu_{i;j}p_iu_{i+1}=\mu_{i;j}p_i(h_i-h_{i+1})=\mu_{i;j}(p_ih_i-p_ih_{i+1})=\mu_{i;j}(p_ih_i+q_ih_{i-1}-h_i),
\end{equation}
the last equality due to (\ref{aux11}), by isolating $p_i$ (we omit the arguments for simplicity). Then
\begin{equation}
\mu_{i;j}p_iu_{i+1}=\mu_{i;j}(h_i(p_i-1)+q_ih_{i-1})=\mu_{i;j}(-h_iq_i+q_ih_{i-1})=\mu_{i;j}q_i(h_{i-1}-h_i)=\mu_{i;j}q_iu_i
\end{equation}
Hence $\mu_{i;j}p_iu_{i+1}=\mu_{i;j}q_iu_i$ and so $p_iu_{i+1}=q_iu_i$. Therefore,
\begin{equation}\label{equ1}
u_{i+1}(\eta_j)=\Big(\frac{q_i(\eta_j)}{p_i(\eta_j)}\Big)u_i(\eta_j)=\Big(\frac{q_i(\eta_j)q_{i-1}(\eta_j)\cdots q_1(\eta_j)}{p_i(\eta_j)p_{i-1}(\eta_j)\cdots p_1(\eta_j)}\Big)u_1(\eta_j)=\gamma_i(\eta_j)u_1(\eta_j)
\end{equation}
where
\begin{equation}\label{gammaeq1}
\gamma_i(\eta_j):=\frac{q_i(\eta_j)q_{i-1}(\eta_j)\cdots q_1(\eta_j)}{p_i(\eta_j)p_{i-1}(\eta_j)\cdots p_1(\eta_j)}=\frac{\lambda_{i;j}\lambda_{i-1;j}\cdots \lambda_{1;j}}{\mu_{i;j}\mu_{i-1;j}\cdots \mu_{1;j}},
\end{equation}
the last equality due to the eigenvalue conditions. Then note that $u_1+\cdots+u_i=h_0^A-h_i^A$ and recalling that $h_0^A=1$ we can write
$$h_i^A=h_0^A-u_1-u_2-\cdots - u_i=1-\gamma_0u_1-\gamma_1u_1-\cdots -\gamma_{i-1}u_1$$
\begin{equation}
=1-B(\gamma_0+\cdots+\gamma_{i-1}),\;\;\;A=u_1,\;\;\;\gamma_0=1
\end{equation}
Here $B$ is yet to be determined. If $\sum_{i=0}^\infty\gamma_i(\eta_j)=\infty$ the restriction $0\leq h_i^A\leq 1$ forces $B=0$ and $h_i^A=1$ for all $i$. If $\sum_{i=0}^\infty\gamma_i(\eta_j)<\infty$ then we may take $B>0$ such that
\begin{equation}
1-B(\gamma_0+\cdots+\gamma_{i-1})\geq 0,\;\;\;\forall i
\end{equation}
So the minimal nonnegative solution occurs when $B=B(j)=(\sum_{k=0}^\infty\gamma_k(\eta_j))^{-1}$ and then for all $j$,
\begin{equation}
h_i^A(\eta_j)=\frac{\sum_{k=i}^\infty\gamma_k(\eta_j)}{\sum_{k=0}^\infty \gamma_k(\eta_j)},\;\;\;\forall i=1,2,\dots
\end{equation}

\qed

{\bf Proof of Theorem \ref{teo_pot11}.} The proof is inspired by a classical description \cite{norris}. Item a) was proved before the statement of this theorem. b) Consider the expected cost up to time $n$:
\begin{equation}\label{cost_upton}
\phi_i(\rho;n):=\sum_{X_0=i,X_1,X_2,\dots\in(-a,a), X_r\in\{-a,a\},r\leq n}[c(X_0)+c(X_1)+c(X_2)+\cdots+c(X_{r-1})+f(X_r)]P_i(X_1,\dots,X_r;\rho),\;\;\;n\geq 1
\end{equation}
and we define $\phi_i(\rho;0)=0$. Then it should be clear that $\phi_i(\rho;n)\uparrow \phi_i(\rho)$ as $n\to\infty$. Similarly,
\begin{equation}
\phi_i(\rho;n+1)=c(i)+\sum_{j\in I}p_{ij}\phi_j\Big(\frac{B_i^j\rho B_i^{j*}}{tr(B_i^j\rho B_i^{j*})};n\Big),\;\;\;n\geq 0
\end{equation}
Now suppose $\psi\geq 0$ satisfies
\begin{equation}\label{qver_desig}
\psi_i(\rho)\geq c(i)+\sum_j p_{ij}(\rho)\psi_j\Big(\frac{B_i^j\rho B_i^{j*}}{tr(B_i^j\rho B_i^{j*})}\Big),
\end{equation}
in $D$ and $\psi_i(\rho)\geq f(i)$ in $\partial D$. Correspondingly, we call (\ref{qver_desig}) the {\bf open quantum version} of $\psi\geq c+P\psi$. Note that for all $\rho$, $\psi_i(\rho)\geq 0=\phi_i(\rho;0)$ so
\begin{equation}
\psi_i(\rho)\geq c(i)+\sum_j p_{ij}(\rho)\psi_j\Big(\frac{B_i^j\rho B_i^{j*}}{tr(B_i^j\rho B_i^{j*})}\Big)\geq c(i)+\sum_j p_{ij}(\rho)\phi_j\Big(\frac{B_i^j\rho B_i^{j*}}{tr(B_i^j\rho B_i^{j*})};0\Big)=\phi_i(\rho;1)
\end{equation}
in $D$. By induction we conclude $\psi_i(\rho)\geq\phi_i(\rho;n)$ and hence $\psi_i(\rho)\geq \phi_i(\rho)$ for all $\rho$, all $i$. For simplicity we write $\psi\geq \phi$.

\medskip

c) Now we assume $P_i(T<\infty)=1$ for all $i$, i.e., we assume the probability of ever reaching $\partial D$ equals 1. Note that a choice of initial density $\rho$ is implicit here. We would like to show that the open quantum version  of the problem $\phi=c+P\phi$ in $D$, $\phi=f$ in $\partial D$ has at most one bounded solution. Let $\psi_i(\rho)$ be another solution. For $i\in D$ we have, writing $p_{ij}=p_{ij}(\rho)=tr(B_i^j\rho B_i^{j*})$ for simplicity,
\begin{equation}
\psi_i(\rho)= c(i)+\sum_{j\in I}p_{ij}\psi_j\Big(\frac{B_i^j\rho B_i^{j*}}{tr(B_i^j\rho B_i^{j*})}\Big)
=c(i)+\sum_{j\in \partial D}p_{ij}f(j)+\sum_{j\in D}p_{ij}\psi_j\Big(\frac{B_i^j\rho B_i^{j*}}{tr(B_i^j\rho B_i^{j*})}\Big)
\end{equation}
By performing a repeated substitution we get
\begin{equation}\label{norris_v1}
\psi_i(\rho)= c(i)+\sum_{j\in \partial D}p_{ij}(\rho)f(j)+\sum_{j\in D}p_{ij}(\rho)\psi_j\Big(\frac{B_i^j\rho B_i^{j*}}{tr(B_i^j\rho B_i^{j*})}\Big)=
\end{equation}
$$=c(i)+\sum_{j\in \partial D}p_{ij}(\rho)f(j)+\sum_{j\in D}p_{ij}(\rho)\Bigg[ c(j)+\sum_{j_1\in \partial D}p_{jj_1}\Big(\frac{B_i^j\rho B_i^{j*}}{tr(B_i^j\rho B_i^{j*})}\Big)f(j_1)+$$
$$+\sum_{j_1\in D}p_{jj_1}\Big(\frac{B_i^j\rho B_i^{j*}}{tr(B_i^j\rho B_i^{j*})}\Big)\psi_{j_1}\Big(\frac{B_j^{j_1}\frac{B_i^j\rho B_i^{j*}}{tr(B_i^j\rho B_i^{j*})} B_j^{j_1*}}{tr(B_j^{j_1}\frac{B_i^j\rho B_i^{j*}}{tr(B_i^j\rho B_i^{j*})} B_j^{j_1*})}\Big) \Bigg]$$
$$
=c(i)+\sum_{j\in \partial D}p_{ij}(\rho)f(j)+\sum_{j\in D}p_{ij}(\rho)\Bigg[ c(j)+\sum_{j_1\in \partial D}p_{jj_1}\Big(\frac{B_i^j\rho B_i^{j*}}{tr(B_i^j\rho B_i^{j*})}\Big)f(j_1)+$$
$$+\sum_{j_1\in D}p_{jj_1}\Big(\frac{B_i^j\rho B_i^{j*}}{tr(B_i^j\rho B_i^{j*})}\Big)\psi_{j_1}\Big(\frac{B_j^{j_1}B_i^j\rho B_i^{j*} B_j^{j_1*}}{tr(B_j^{j_1}B_i^j\rho B_i^{j*} B_j^{j_1*})}\Big) \Bigg]
$$
$$=c(i)+\sum_{j\in \partial D}p_{ij}(\rho)f(j)+\sum_{j\in D}p_{ij}(\rho)c(j)+\sum_{j\in D}\sum_{j_1\in \partial D}p_{ij}(\rho)p_{jj_1}\Big(\frac{B_i^j\rho B_i^{j*}}{tr(B_i^j\rho B_i^{j*})}\Big)f(j_1)\;+
$$
\begin{equation}\label{norris_v2}
+\sum_{j\in D}\sum_{j_1\in D}p_{ij}(\rho)p_{jj_1}\Big(\frac{B_i^j\rho B_i^{j*}}{tr(B_i^j\rho B_i^{j*})}\Big)\psi_{j_1}\Big(\frac{B_j^{j_1}B_i^j\rho B_i^{j*} B_j^{j_1*}}{tr(B_j^{j_1}B_i^j\rho B_i^{j*} B_j^{j_1*})}\Big)=\cdots
\end{equation}
$$\cdots=c(i)+\sum_{j\in\partial D} p_{ij}(\rho)f(j)+\sum_{j\in D} p_{ij}(\rho)c(j)$$
$$
+\cdots+\sum_{j_1\in D}\cdots\sum_{j_{n-1}\in D} p_{ij}p_{jj_1}\cdots p_{j_{n-2}j_{n-1}}c(j_{n-1})+\sum_{j_1\in D}\cdots\sum_{j_{n-1}\in D}\sum_{j_n\in\partial D} p_{ij}p_{jj_1}\cdots p_{j_{n-1}j_n}f(j_n)$$
\begin{equation}\label{norris_v3}
+\sum_{j_1\in D}\cdots\sum_{j_n\in D} p_{ij_1}p_{jj_1}\cdots p_{j_{n-1}j_n}\psi_{j_n}
\end{equation}
Note that in the last equality we have omitted the dependence of the $p_{ij}$ on the density matrices. A convenient cancellation of terms occur in each of them. For instance, in the last term, $p_{ij_1}\cdots p_{j_{n-1}j_n}$ simplifies to
\begin{equation}
p_{ij_1}\cdots p_{j_{n-1}j_n}=tr(B_i^{j_1}\rho B_i^{j_1*})tr\Big(\frac{B_{j_1}^{j_2}B_i^{j_1}\rho B_i^{j_1*}B_{j_1}^{j_2*}}{tr(B_i^{j_1}\rho B_i^{j_1*})}\Big)\cdots=tr(B_{j_{n-1}}^{j_n}\cdots B_i^{j_1}\rho B_i^{j_1*}\cdots B_{j_{n-1}}^{j_n*})
\end{equation}

Now suppose $P_i(T<\infty)=1$ for all $i$ and that $|\psi_i|\leq M$ then as $n\to \infty$,
\begin{equation}
\Big|\sum_{j_1\in D}\cdots\sum_{j_n\in D} p_{ij_1}\cdots p_{j_{n-1}j_n}\psi_{j_n}\Big|\leq MP_i(T\geq n)\to 0,
\end{equation}
which means the last term in (\ref{norris_v3}) vanishes when $n\to\infty$. Recalling definition (\ref{cost_upton}) we obtain in the case of equality in eqs. (\ref{norris_v1})-(\ref{norris_v3}) that $\psi_i=\lim_{n\to\infty}\phi_i(\rho;n)=\phi_i$.

\qed


\begin{thebibliography}{99}
\bibitem{accardi1} L. Accardi, D. Koroliuk. Stopping times for quantum Markov chains. Journ. Theor. Probability, Vol. 5, no. 3, pp 521-535, 1992.
\bibitem{accardi2} L. Accardi, D. Koroliuk. Quantum Markov chains: The recurrence problem. Quantum Prob. and Related Topics VII, 63–73 (1991).
\bibitem{alicki} R. Alicki, M. Fannes. Quantum Dynamical Systems. Oxford University Press, Oxford, 2000.

\bibitem{attal} S. Attal, F. Petruccione, C. Sabot, I. Sinayskiy. Open Quantum Random Walks. J. Stat. Phys. (2012) 147:832852.
\bibitem{attal2} S. Attal, N. Guillotin-Plantard, C. Sabot. Central Limit Theorems for Open Quantum Random Walks and Quantum Measurement Records. Ann. Henri Poincar\'e 16 (2015), 15-43.
\bibitem{bllt2} A. Baraviera, C. F. Lardizabal, A. O. Lopes, M. Terra Cunha. Quantum stochastic processes, quantum iterated function systems and entropy. S\~ao Paulo Journ. Math. Sci., Vol. 5, N 1., p. 51-84, 2011.
\bibitem{benatti} F. Benatti. Dynamics, Information and Complexity in Quantum Systems. Springer, Berlin (2009).
\bibitem{bremaud} P. Br\'emaud. Markov Chains: Gibbs Fields, Monte Carlo Simulation and Queues. Texts in Applied Mathematics 31. Springer, 1999.
\bibitem{konno2} P. Biane, L. Bouten, F. Cipriani, N. Konno, N. Privault, Q. Xu. Quantum Potential Theory. Lecture Notes in Mathematics 1954 Springer-Verlag Berlin Heidelberg (2008).
\bibitem{bourg} J. Bourgain, F. A. Gr\"unbaum, L. Vel\'azquez, J. Wilkening. Quantum recurrence of a subspace and operator-valued Schur functions, Comm. Math. Phys., 329 (2014) 1031-1067.
\bibitem{bratteli} O. Bratteli, D. W. Robinson. Operator Algebras and Quantum Statistical Mechanics I. Springer-Verlag, 1979.
\bibitem{burgarth} D. Burgarth, G. Chiribella, V. Giovannetti, P. Perinotti, K. Yuasa. Ergodic and Mixing Quantum Channels in Finite Dimensions. 	 New J. Phys. 15, 073045 (2013).
\bibitem{burgarth2} D. Burgarth, V. Giovannetti. The generalized Lyapunov theorem and its application to quantum channels. New Journal of Physics 9 (2007) 150.
\bibitem{carbone} R. Carbone, Y. Pautrat. Homogeneous open quantum random walks on a lattice. Preprint arXiv:1408.1113v2.
\bibitem{carbone2} R. Carbone, Y. Pautrat. Open quantum random walks: reducibility, period, ergodic properties. Preprint arXiv:1405.2214v3.
\bibitem{fagnola} F. Fagnola, R. Rebolledo. Transience and recurrence of quantum Markov semigroups. Probab. Theory Relat. Fields 126, 289–306 (2003).
\bibitem{grimmett} G. R. Grimmett, D. R. Stirzaker. Probability and Random Processes, 3rd edition. Oxford University Press, 2001.
\bibitem{gudder2} S. Gudder. Transition effect matrices and quantum Markov chains. Found. Phys. (2009) 39:573-592.
\bibitem{gudder} S. Gudder. Quantum Markov chains. Journ. Math. Phys. 49, 072105 (2008).
\bibitem{werner} F. A. Gr\"unbaum, L. Vel\'azquez, A. H. Werner, R. F. Werner. Recurrence for Discrete Time Unitary Evolutions. Comm. Math. Phys. 320, 543–569 (2013).
\bibitem{hj1} R. A. Horn, C. R. Johnson. Matrix analysis. Cambridge University Press, 1985.
\bibitem{hj2} R. A. Horn, C. R. Johnson. Topics in matrix analysis. Cambridge University Press, 1991.
\bibitem{konno} N. Konno, H. J. Yoo. Limit Theorems for Open Quantum Random Walks. J. Stat. Phys. (2013) 150:299-319.
\bibitem{maassen} B. K\"ummerer, H. Maassen. A pathwise ergodic theorem for quantum trajectories. J. Phys. A Math. Gen. 37, 11889-11896 (2004).
\bibitem{cfrr} C. F. Lardizabal, R. R. Souza. On a class of quantum channels, open random walks and recurrence. J. Stat. Phys. (2015) 159:772-796.
\bibitem{lardi} C. F. Lardizabal. A quantization procedure based on completely positive maps and Markov operators. Quantum Inf. Process. (2013) 12:1033-1051.
\bibitem{liu} C. Liu. From open quantum walks to unitary quantum walks. arXiv:1502.01680.
\bibitem{petulante} C. Liu, N. Petulante. On Limiting Distributions of Quantum Markov Chains. Int. J. Math. and Math. Sciences. Volume 2011, ID 740816.
\bibitem{petulante2} C. Liu, N. Petulante. Asymptotic evolution of quantum walks on the N-cycle subject to decoherence on both the coin and position degrees of freedom. Phys. Rev. E 81, 031113 (2010).
\bibitem{lozinski} A. Lozinski, K. \.Zyczkowski, W. S\l omczy\'{n}ski. Quantum iterated function systems, Physical Review E, Volume 68, 04610, 2003.
\bibitem{lmei} E. Lytvynov, L. Mei. On the correlation measure of a family of commuting Hermitian operators with applications to particle
densities of the quasi-free representations of the CAR and CCR. Journ. Funct. Analysis 245 (2007) 62–88.
\bibitem{meyer} P.-A. Meyer. Quantum probability for probabilists. Lecture notes in mathematics 1538, Springer, Berlin 1993.
\bibitem{mukhamedov} F. Mukhamedov. Weak ergodicity of nonhomogeneous Markov chains on noncommutative $L^1$-spaces. Banach J. Math. Anal. 7 (2013), no. 2, 53-73.
\bibitem{nielsen} M. A. Nielsen, I. L. Chuang. Quantum computation and quantum information. Cambridge University Press, 2000.
\bibitem{norris} J. R. Norris. Markov chains. Cambridge University Press, 1997.
\bibitem{novotny} J. Novotn\'y, G. Alber, I. Jex. Asymptotic evolution of random unitary operations. Cent. Eur. J. Phys. 8(6), 1001-1014, 2010.
\bibitem{novotny2} J. Novotn\'y, G. Alber, I. Jex. Asymptotic properties of quantum Markov chains. J. Phys. A: Math. Theor. 45 485301 (2012).
\bibitem{oliveira} C. R. de Oliveira. Intermediate spectral theory and quantum dynamics. Birkh\"auser Verlag 2009.
\bibitem{partha} K. R. Parthasarathy. An introduction to quantum stochastic calculus. Birkh\"auser, 1992.
\bibitem{paulsen} V. Paulsen. Completely Bounded Maps and Operator Algebras, Cambridge University Press, 2003.
\bibitem{pellegrini} C. Pellegrini. Continuous Time Open Quantum Random Walks and Non-Markovian Lindblad Master Equations. J. Stat. Phys. (2014) 154:838-865.
\bibitem{petz} D. Petz. Quantum Information Theory and Quantum Statistics. Springer, 2008.
\bibitem{platis} A. Platis, N. Limnios, M. Le Du. Hitting time in a finite non-homogeneous Markov chain with applications. Appl. Stochastic Models \& Data Anal., 14, 241-253 (1998).
\bibitem{portugal} R. Portugal. Quantum walks and search algorithms. Springer, 2013.
\bibitem{ruskaisw} C. King, M. B. Ruskai. Minimal Entropy of States Emerging from Noisy Quantum Channels. IEEE Trans. Info. Theory, 47, 192-209 (2001).
\bibitem{salvador} S. E. Venegas-Andraca. Quantum walks: a comprehensive review. Quantum Inf. Process. (2012) 11:1015-1106.
\bibitem{saloff} L. Saloff-Coste, J. Z\'u\~niga. Convergence of some time inhomogeneous Markov chains via spectral techniques. Stoch. proc. and their applications. 117 (2007) 961-979.
\bibitem{szehr} O. Szehr, M. M. Wolf. Perturbation bounds for quantum Markov processes and their fixed points. J. Math. Phys. 54, 032203 (2013).
\bibitem{temme} K. Temme, M. J. Kastoryano, M. B. Ruskai, M. M. Wolf, F. Verstraete. The $\chi^2$-divergence and mixing times of quantum Markov processes. J. Math. Phys. 51, 12201 (2010).
\bibitem{watrous} J. Watrous. Theory of Quantum Information Lecture Notes from Fall 2011. Institute for Quantum Computing, University of Waterloo.
\bibitem{wolf} M. M. Wolf. Quantum Channels \& Operations Guided Tour Lecture Notes (unpublished). July 5, 2012.
\end{thebibliography}
\end{document}